\documentclass[11pt,a4paper]{article}
\pdfoutput=1
\usepackage{jheppub}
\usepackage{rotating}
\usepackage{array}
\usepackage{amsmath}
\usepackage[normalem]{ulem}
\usepackage{cancel}
\usepackage{slashed}
\usepackage{booktabs}
\usepackage[pdftex,table]{xcolor}
\usepackage{units}
\usepackage{xspace}
\usepackage{subcaption}
\usepackage{graphicx}
\usepackage{algorithm}
\usepackage{algpseudocode}
\usepackage{rotating}
\usepackage{adjustbox}
\usepackage{tabularx}
\usepackage{bm}
\usepackage{rotating}
\usepackage{caption}
\usepackage{todonotes}
\usepackage{hyperref}

\newcommand{\ba}[1]{\ensuremath{\left( #1 \right)}}
\newcommand{\bb}[1]{\ensuremath{\left[ #1 \right]}}
\newcommand{\bc}[1]{\ensuremath{\left\{ #1 \right\}}}

\newcommand{\td}[2]{\ensuremath{\frac{\mathrm d #1}{\mathrm d #2}}}
\newcommand{\tdd}[2]{\ensuremath{\frac{\mathrm d^2 #1}{\mathrm d #2^2}}}

\newcommand*\diff{\mathop{}\!\mathrm{d}}

\newcommand{\DS}{\ensuremath{\text{DS}}}
\newcommand{\SM}{\ensuremath{\text{SM}}}

\newcommand{\beq}{\begin{equation}}
\newcommand{\eeq}{\end{equation}}
\newcommand{\dd}{\ensuremath{\text{d}}}

\definecolor{DESYcyan}{RGB}{0,159,223}
\definecolor{DESYorange}{RGB}{241,143,31}
\definecolor{DESYrot}{RGB}{235,90,45}
\definecolor{DESYdunkelrot}{RGB}{185,45,65}
\definecolor{DESYdunkelblau}{RGB}{0,75,110}
\definecolor{DESYviolett}{RGB}{145,125,185}
\definecolor{DESYlila}{RGB}{80,80,155}
\definecolor{DESYgelb}{RGB}{250,200,0}

\definecolor{PlotGreen}{HTML}{3cb44b}
\definecolor{PlotRed}{HTML}{e6194b}
\definecolor{PlotYellow}{RGB}{255, 225, 25}

\usepackage{multirow}

\arxivnumber{DESY-26-014, P3H-26-009, TTP26-002}

\title{Tuning the violins:~dark sector phase transition models for the PTA signal}

\author[1]{Torsten Bringmann,}
\author[2]{Thomas Konstandin,}
\author[3]{Jonas Matuszak,}
\author[2,4]{Kai Schmidt-Hoberg,}
\author[5,6]{and Carlo Tasillo}

\affiliation[1]{Department of Physics, University of Oslo, Box 1048, N-0316 Oslo, Norway}
\affiliation[2]{Deutsches Elektronen-Synchrotron DESY, Notkestr.~85, 22607 Hamburg, Germany}
\affiliation[3]{Institute for Theoretical Particle Physics (TTP), Karlsruhe Institute of
  Technology (KIT), 76128 Karlsruhe, Germany}
\affiliation[4]{CP3-Origins, University of Southern Denmark,
Campusvej 55, DK-5230 Odense M, Denmark}
\affiliation[5]{Department of Physics and Astronomy, Uppsala University, Box 516, SE-751 20 Uppsala, Sweden}
\affiliation[6]{Instituto de Física Corpuscular (IFIC), Universitat de Valéncia-CSIC, Parc Científic UV,
\\C/ Catedrático José Beltrán 2, E-46980 Paterna, Spain}

\emailAdd{torsten.bringmann@fys.uio.no}
\emailAdd{thomas.konstandin@desy.de}
\emailAdd{jonas.matuszak@kit.edu}
\emailAdd{kai.schmidt-hoberg@desy.de}
\emailAdd{carlo.tasillo@ific.uv.es}

\abstract{
First-order phase transitions in a dark sector have been invoked as 
an intriguing possibility to explain the observed stochastic gravitational
wave background at nanohertz frequencies. 
Here we perform a comprehensive study of the generic requirements for 
such a phase transition to 
explain the observed signal while being consistent with all relevant
constraints. We consider three broad model classes for strong first-order transitions,
realised by an Abelian dark Higgs boson, a two-step phase transition involving two
scalar singlets, and a conformal scalar field with loop-induced symmetry breaking,
respectively. We discuss the 
tuning that is required to successfully explain the Pulsar Timing Array (PTA) signal in each of these cases, and 
highlight the underlying physical mechanisms. 
We conclude that all three scenarios can in principle describe the data, but that conformal models 
stand out as the most generic, and least tuned, explanation.
Future observations by the PTA collaborations and collider experiments will be crucial
to test the viability of this hypothesis, and to further narrow in on the model parameters,
if the PTA signal is indeed due to a strong first-order phase transition.
}

\keywords{Phase Transitions in the Early Universe, Early Universe Particle Physics, Cosmology of Theories BSM}

\begin{document}
\maketitle
\flushbottom

\section{Introduction}\label{sec:intro}

The first observation of gravitational waves (GWs) in 2015 -- originating from a binary black hole merger at 
frequencies of $\mathcal{O}(100)$~Hz~\cite{LIGOScientific:2016aoc} -- marked a scientific breakthrough and opened a 
new window into the Universe.
Depending on the astrophysical or cosmological source, gravitational waves can be
produced across many orders of magnitude in frequency, motivating a range of complementary experimental strategies~\cite{LISA:2017pwj, NANOGrav:2020bcs}. 
Pulsar timing arrays (PTAs), in particular, are sensitive to very 
small frequencies in the nanohertz (nHz) regime. In 2020, the NANOGrav collaboration reported evidence for a common 
red-noise process suggestive of a stochastic nHz gravitational-wave background (GWB)~\cite{NANOGrav:2020bcs}. 
Subsequent data releases from NANOGrav~\cite{NANOGrav:2023gor},  EPTA \& InPTA~\cite{Antoniadis:2023ott}, 
PPTA~\cite{Reardon:2023gzh}, CPTA~\cite{Xu:2023wog}, and MPTA~\cite{Miles:2024seg} confirmed the presence 
of this signal and also displayed increasing evidence for the characteristic Hellings-Downs 
correlation associated with a GWB~\cite{Hellings1983}. It is possible that the discovery threshold will already be passed within the upcoming third data 
release of the International Pulsar Timing Array (IPTA)~\cite{InternationalPulsarTimingArray:2023mzf, Yu:2025dor}. This development 
makes it both compelling and timely to explore potential origins of the GWB signal.

Supermassive black hole binaries (SMBHBs) are astrophysical sources known to contribute to the GWB in this 
frequency range~\cite{NANOGrav:2020bcs}. However, to match the observed signal amplitude the local SMBHB 
number density, merger rates and masses would need to be significantly larger than previously 
estimated~\cite{Casey-Clyde2021,Kelley2016,Kelley2017,NANOGrav:2023hfp, Chen:2025wel}; also the predicted
power-law spectrum is in some tension with the observed signal shape~\cite{Agazie:2026tui}. In addition, it has been pointed out that 
the mechanism by which SMBHBs dissipate sufficient energy to come close enough for relevant emission of a GWB
is not yet fully understood,  a conundrum coined as the 
`final parsec problem'~\cite{Milosavljevic:2002ht,Chiaberge:2025ouh}. 
It is therefore interesting to consider additional sources of a nHz GWB and it has been widely recognised that there is 
the exciting possibility that the observed signal may originate from genuinely cosmological 
sources~\cite{NANOGrav:2023hvm} such as inflation~\cite{Vagnozzi:2020gtf}, topological 
defects~\cite{Blasi2020,Ellis2020b,Buchmuller:2020lbh,Samanta:2020cdk,Guo:2023hyp,King:2023cgv},
the generation of sub-solar-mass primordial black holes (PBHs)\footnote{%
There is also the possibility that merging primordial SMBHs are responsible for the signal~\cite{Depta:2023uhy}.
}~\cite{DeLuca:2020agl,Vaskonen:2020lbd,Kohri:2020qqd, Unal:2020mts,Ashoorioon:2022raz} 
or first-order phase transitions 
(FOPTs)~\cite{Nakai2020,Ratzinger:2020koh,NANOGrav:2021flc,Bringmann:2023opz,Madge:2023dxc,Goncalves:2025uwh}.  Often these 
scenarios are specifically tailored to explain the PTA data and it is not always clear how `natural' or realistic the 
underlying models are.

In this work we will concentrate on the discussion of FOPTs and focus in particular on the question of what 
is required from the underlying particle physics point of view. 
The generic requirements on a FOPT to fit the data, while at the 
same time respecting cosmological constraints, has been laid out in detail in ref.~\cite{Bringmann:2023opz}. 
Specifically, a phase transition  capable of generating gravitational waves at nanohertz frequencies must occur at 
temperatures around the MeV scale. To fit the data the phase transition  should furthermore be strong and proceed extremely 
slowly, resulting in almost Hubble-sized bubbles at percolation. However, new physics at such low energies is already 
tightly constrained by numerous direct searches~\cite{Gori:2022vri}, which strongly suggests that any such phase 
transition would have to occur within a largely 
secluded dark sector. Despite weak couplings to the SM, a dark sector phase transition (DSPT) can still leave 
observable imprints. In particular, the additional energy density stored in the dark sector can modify the predictions of 
Big Bang Nucleosynthesis (BBN) and affect the Cosmic Microwave Background (CMB) 
anisotropies~\cite{Planck:2018vyg,Yeh:2022heq,Bagherian:2025puf}. 
Late decays of dark sector states may also 
lead to conflicts with BBN or CMB 
observations~\cite{Hufnagel:2018bjp, Forestell:2018txr,Depta:2020zbh,Depta:2020mhj,Kawasaki:2020qxm}. 
Taking into account these constraints, a DSPT could potentially explain the observed signal if (and only if)  
the dark sector states decay before the time of neutrino decoupling~\cite{Bringmann:2023opz}.

Here we update the model-independent analysis from ref.~\cite{Bringmann:2023opz} to take into account the most 
recent NANOGrav 15~yr data set, 
finding that the preferred phase transition  parameters shift towards larger temperatures (which generally relaxes 
constraints from BBN). We 
study three underlying particle physics models in detail, which are representative of the 
different qualitative possibilities to satisfy the above requirements, with respectively a
\begin{enumerate}
\item thermally induced barrier (as realised in a model with a dark Abelian Higgs boson)
\item two-step phase transition (as realised in a ``flip-flop'' model,  with two scalar singlets)
\item loop-induced barrier (as realised in the conformal version of the dark $U(1)'$ model).
\end{enumerate}
All these model classes allow a priori for a \emph{strong} FOPT. 
In all cases, we evaluate in detail what is required for a good fit to the data while simultaneously 
being in agreement with cosmological and other complementary constraints.
Our main conclusion is that while on a model-independent level a FOPT seems to be a generically 
viable explanation, this conclusion is not generic once the microscopic physics is taken into account.
Instead, for both the cases of \emph{(i)} the Abelian dark Higgs and \emph{(ii)} the two-stage 
phase transition with two singlets, significant parameter tuning is necessary to fit the data. 
In contrast, for models that are (nearly) conformal, the required parameter space can be realised much 
more naturally, making this class particularly appealing from a model-building perspective. 

This work is organised as follows: In section~\ref{sec:model-independent}, we present an updated model-indepen\-dent
analysis of the NANOGrav 15\,yr data in terms of a DSPT. In section~\ref{sec:models}, we discuss the three
representative particle physics models in detail. In section~\ref{sec:results}, we confront the models with the PTA data and
quantify the required tuning to satisfy all constraints. Section~\ref{sec:discussion} contains a discussion of our results
and we conclude in section~\ref{sec:conclusion}.
In three appendices, we provide further details about our PTA 
likelihood implementation (\ref{app:ptarcade}), the flip-flop model (\ref{app:flipflop}), 
as well as best-fit points and preferred model parameter ranges for each of the model classes (\ref{app:triangles}).
 
\section{Model-independent analysis}
\label{sec:model-independent}

In our analysis we will consider the possibility that the GW background at nHz frequencies is dominantly
produced by a FOPT in a dark sector. For simplicity, we will typically assume the dark sector to be in thermal 
equilibrium with the SM bath throughout the transition, but we will also discuss implications of deviating
from this assumption.
For the scenarios that we are interested in, there are three 
main phenomenological parameters that determine the shape of the GW spectrum: the transition strength parameter $\alpha$,
quantifying the amount of vacuum energy released in the transition; the inverse timescale $\beta/H$ of the transition, 
determined by the mean bubble separation $RH_*$ at percolation (both typically stated in units of
the Hubble parameter $H$ at percolation); and the reheating temperature $T_\text{reh}$, which sets 
the scale and hence the peak frequency of the GW spectrum. In general, also the bubble wall velocity $v_\text{w}$ 
impacts both overall spectral shape and peak position of the GW spectrum, but we find $v_\text{w}\approx1$ in almost 
all cases of relevance discussed below.

In this section, we will keep the analysis largely model-independent and point out the generic requirements on these
phenomenological parameters that must be met in order to provide a good fit to the PTA data. In section~\ref{sec:models},
we will then turn to the model-building perspective that is the main focus of this article, by discussing the range of 
dark sector models that could potentially meet the generic requirements outlined here. 

Concretely, we will for simplicity consider a situation where the GW production from the DSPT is dominated by sound
waves, motivated by the already mentioned fact that we will typically encounter non-runaway bubbles. 
For these contributions we follow ref.~\cite{Caprini:2024hue} and use the 
simulation-based~\cite{Jinno:2022mie,Caprini:2024gyk} template
\begin{align}
    \label{eq:GWspec}
    h^2 \Omega_\text{gw}^\text{PT}(f) = \mathcal{R}h^2 A_{\mathrm{sw}} K^2 \mathcal{Y}_{\mathrm{sw}} \ba{RH_*}  \tilde{S}(f)\,,
\end{align}
for the gravitational wave spectrum. This spectrum takes the form of a doubly broken 
power law with spectral index 3 at small frequencies ($f\ll f_1<f_2$) and spectral index $-3$ at large frequencies ($f\gg f_2$):
\begin{align}
  \tilde S(f) &= N {\left( \frac{f}{f_2} \right)}^{3} {\left[ 1 + {\left( \frac{f}{f_1} \right)}^2 \right]}^{-1}
  \left[ 1 + \left( \frac{f}{f_2} \right)^4 \right]^{-1},
\end{align}
with normalisation $N=\frac{2\sqrt{2}}{\pi}\left[(1+f_2^2/f_1^2)+\sqrt{2}f_2/f_1\right]$.
The two frequency breaks occur at
\begin{align}\label{eq:fbreaks}
  f_1 &\simeq  0.2 \ba{\frac{a_*  H_{*}}{RH_*} }\qquad \mathrm{and} \qquad f_2 \simeq 0.5\, \Delta_{\text{w}}^{-1} \ba{\frac{a_* H_{*}}{RH_*} }.
\end{align}
Here, $\Delta_\text{w} \equiv \left| v_{\text{w}} - c_{\text{s}} \right|/\max(v_{\text{w}}, c_{\text{s}})$ parametrizes 
the sound shell thickness, with $c_\text{s}$ being the sound speed in the  (broken phase) plasma. 
The mean bubble separation $RH_*$ at percolation
relates to the transition speed as
\begin{align}
\label{eq:bH}
\beta/H = \left(\frac{8 \pi}{1 - P_\text{f}(T_*)}\right)^{\frac13}\,\frac{\max (c_\text{s}, v_\text{w})}{RH_*}\,,
\end{align}
where $P_\text{f}(T_*) \approx 0.71$
is the false vacuum fraction at percolation~\cite{TL-release, Megevand:2016lpr}. 
The red-shifted Hubble rate, finally,  is given by
\begin{align}\label{eq:Hredshift}
  a_* H_{*} &=  \, 11.2 \, \text{nHz} \,\ba{\frac{T_\text{reh}}{100 \, \text{MeV}}}  \ba{\frac{g_{\text{reh}}}{10} }^{1/2}
             \ba{\frac{10}{h_{\text{reh}}}}^{1/3}\,,
\end{align}
with $g_{\text{reh}}$ ($h_{\text{reh}})$ denoting the combined effective number of
energy (entropy) degrees of freedom at reheating,  in the thermalised SM and dark sector bath.
For large values of $\alpha$, we expect $v_\text{w} \approx 1$ and hence $f_1\sim f_2$, i.e.~almost resulting in 
a power-law with a single break. For small $\alpha$, on the other hand, this is not necessarily the case and
the plateau-like intermediate region has to be taken into account for a sufficiently accurate modelling of the 
GW spectrum; in terms of the concrete models that we discuss further down, this is however only relevant in a small part 
of the flip-flop model parameter space. 

The overall normalization of the spectrum in eq.~\eqref{eq:GWspec} is directly taken from simulations, with
$A_{\mathrm{sw}} \approx 0.11$, and then further redshifted by a factor of
\begin{align}
  \mathcal{R}h^2 = \ba{\frac{a_\text{reh}}{a_0}}^4 \ba{\frac{H_\text{reh}}{H_0}}^2 h^2
  = \Omega_{\gamma}h^2 \ba{\frac{h_{0}}{h_{\text{reh}}}}^{4/3} \frac{g_\text{reh}}{g_\gamma} \, ,
\end{align}
where $\Omega_\gamma h^2 = 2.473 \cdot 10^{-5}$ is the present  energy
density in radiation~\cite{Planck:2018vyg}, 
$g_\gamma = 2$ and $h_{0} = 3.91$.\footnote{%
For definiteness, we will assume $g_\DS = 2$ light dark sector degrees of freedom in this section, 
i.e.~$g_\text{tot}^\text{reh} = g_\SM(T_\SM^\text{reh}) + 2$ and 
$h_\text{tot}^\text{reh} = h_\SM(T_\SM^\text{reh}) + 2$. We stress that this choice has no noticeable
effect on our results in the $\alpha-\beta/H$ plane because 
a change in $g_\text{DS}$ can always be absorbed by a small shift in $T_\text{reh}$.
} 
The signal is also suppressed by a kinetic energy fraction $K = 0.6 \kappa_\text{sw} \alpha / (1 + \alpha)$, where we 
compute the efficiency factor $\kappa_\text{sw}=\kappa_\text{sw}(\alpha,v_\text{w})$ following 
refs.~\cite{Espinosa:2010hh,Jinno:2022mie}. Finally, we take into account the timescale on which sound waves
contribute to the GW signal by introducing the factor
$\mathcal{Y}_\text{sw} \equiv \text{min}\ba{1, \tau_\text{sh} H_*}$~\cite{Balan:2025uke}, 
with $\tau_\text{sh} H_* \simeq RH_* / \sqrt{3K/4}$.

\begin{figure}[t]
	\centering
   \includegraphics[width=\linewidth]{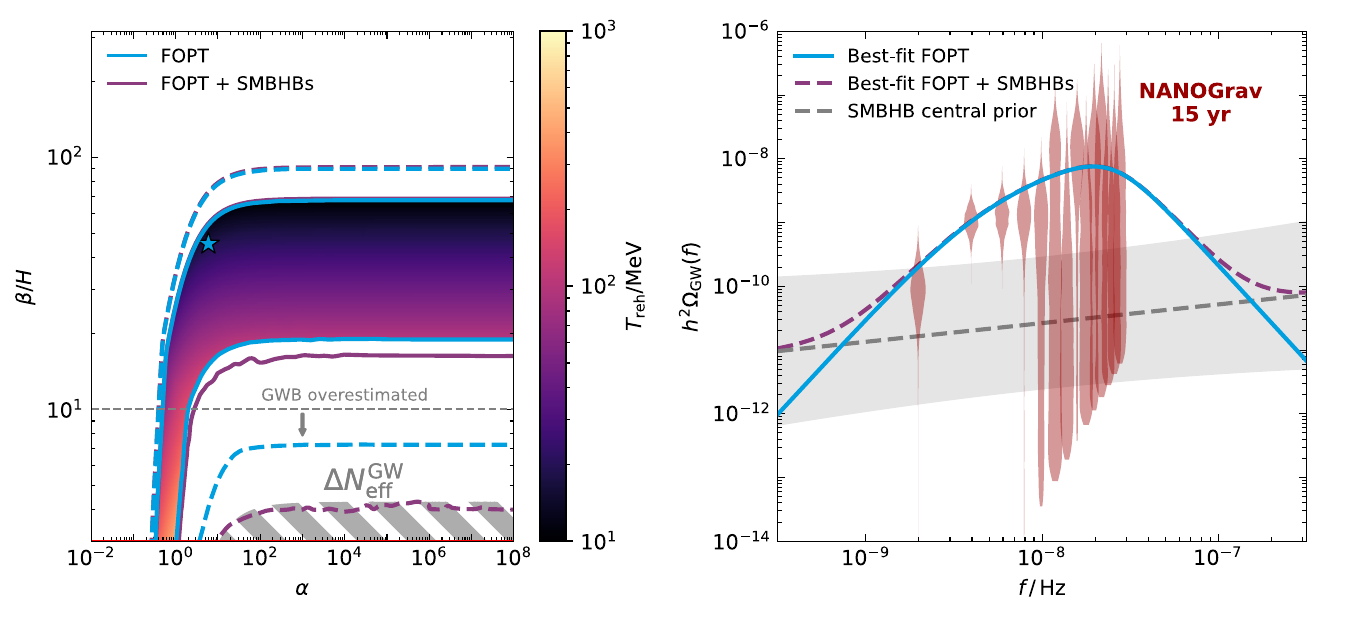}
	\caption{
	 {\it Left panel:}~$1\sigma$ and $2\sigma$ best-fit 
	 regions of current PTA data, interpreted in terms of  strength $\alpha$ and speed $\beta/H$
	of a generic dark sector FOPT (blue lines), with gravitational waves dominantly produced from sound waves due to 
	the expanding bubbles. The colour code shows the corresponding reheating temperature $T_\text{reh}$, and the 
	hatched grey region indicates the combination of parameters that would generate a total energy density in GWs in 
	conflict with $\Delta N_{\rm eff}$ constraints from BBN and CMB~\cite{Yeh:2022heq}. The dashed grey line indicates 
	below which $\beta/H$ values the computation of the GW signal starts to become unreliable~\cite{Jinno:2022mie} 
	as it is not backed up by simulations. The purple lines indicate the $1\sigma$ and 
	$2\sigma$ regions when adding an astrophysical contribution from merging SMBH binaries to the GW signal.
	{\it Right panel:}  Gravitational wave spectra and NANOGrav 15 yr spectrograms~\cite{NANOGrav:2023gor} 
	for the best-fit point when allowing both a phase transition  and merging SMBH contribution (marked with a blue star in the left 
	panel). The grey area indicates the effect of 
	varying amplitude  and slope of the SMBH contribution within their $1\sigma$ 
	expectation~\cite{NANOGrav:2023hvm}.}
	\label{fig:model-independent}
\end{figure}

We now want to compare the generic GW template in eq.~\eqref{eq:GWspec} to the PTA data. For this, we use the 
PTArcade likelihood~\cite{Mitridate:2023oar} of the NANOGrav 15\,yr data set~\cite{NANOGrav:2023gor}, taking into 
account the first 14 Fourier frequencies in the range $f = [2, 28]$\,nHz as recommended by the NANOGrav 
collaboration~\cite{NANOGrav:2023hvm}. 
In the left panel of figure~\ref{fig:model-independent} we show the results of a profile likelihood analysis in the
$\alpha-\beta/H$ plane, i.e.~the iso-likelihood regions after profiling over $T_\text{reh}$ for each point in the plane.
The blue solid (dashed) lines indicate the preferred parameter region at the $1\sigma$ ($2\sigma$) level, respectively. As noted 
previously~\cite{Bringmann:2023opz,NANOGrav:2023hvm, Baldes:2018emh,Levi:2022bzt, Madge:2023dxc,Gouttenoire:2023pxh, Goncalves:2025uwh,
Fujikura:2023lkn,Hosseini:2023qwu,Salvio:2023blb}, a phase transition  interpretation
of the PTA data generally requires low $\beta/H$ and (moderately) large $\alpha$, corresponding to slow and strong  
transitions. The slower the transition, in particular, the less strong it needs to be.

Within the preferred region in the $\alpha-\beta/H$ plane, we also indicate the corresponding reheating 
temperatures $T_{\text{reh}}$ that are needed to explain the data. Ranging from a few $10 \, \text{MeV}$ to a few $100 \, \text{MeV}$,
these are higher temperatures than inferred from the NANOGrav 12.5\,yr data~\cite{Bringmann:2023opz}, which relaxes 
cosmological constraints (see below) significantly. The reason is that the observed signal, see right panel of the figure, 
is now best fit by the rising part of the phase transition  spectrum.  Compared to the peak and UV tail of the GW spectrum, which best 
described the earlier 12.5~yr data release, this corresponds to a preference for larger peak frequencies and hence larger values
of $T_{\text{reh}}$.

Let us briefly discuss the impact of two assumptions that we have made in producing
figure~\ref{fig:model-independent}. First, we checked explicitly that even choosing
smaller bubble wall velocities $v_\text{w}=0.7$, 
rather than $v_\text{w}\approx1$ as for the rest
of our analysis, only has a minor effect in shifting the allowed contour towards slightly
smaller values of $\beta/H$ in the left panel of the figure. The reason is that the resulting
change in normalisation and peak frequency 
can largely be compensated by adjusting the reheating temperature. Another crucial
assumption we have made is thermal equilibrium of dark sector and standard model
throughout the transition. The temperature of a secluded dark sector $T_{\text{DS}}$, however,
may well be different from that of the standard model, $T$. We checked that assuming an
initially hotter dark sector, $\xi\equiv T_{\text{DS}}/T>1$, has the effect of allowing for slightly
smaller values of $\alpha$: instead of $\alpha\gtrsim 0.4$ for $\xi=1$, the left corner of the preferred
region extends down to $\alpha\simeq 0.1$ for $\xi\simeq2$ (while increasing $\xi$ further only has a minimal
effect). For $\xi<1$, on the other hand, the rough $\Omega\propto \xi^4$ scaling (dominantly entering
through $K$~\cite{Ertas:2021xeh}) implies that the small-$\alpha$ solutions visible in the figure will no longer be
viable; for $\xi=0.5$, e.g., the allowed parameter space shown in the left panel of
figure~\ref{fig:model-independent} is restricted to $\alpha\gtrsim 2$.

In the discussion so far we have neglected the expected stochastic background from merging SMBH binaries in
our analysis~\cite{NANOGrav:2023hfp}. In order to quantify the potential impact of this contribution on
our conclusions, we next consider a combined signal of the form
$ \Omega_\text{gw}(f) = \Omega_\text{gw}^\text{PT}(f) + \Omega_\text{gw}^\text{BH}(f)$. Here, we model
$\Omega_\text{gw}^\text{BH}$ as a power-law, with amplitude and slope as nuisance parameters
that are distributed according to the likelihood implemented in
PTArcade~\cite{Mitridate:2023oar} (and recommended by the NANOGrav
collaboration~\cite{NANOGrav:2023hvm}). We show the $1\sigma$ and $2\sigma$ regions
resulting from profiling over both $T_\text{reh}$ and the SMBH nuisance parameters as
purple lines in the left panel of figure~\ref{fig:model-independent}. Consistent with
earlier findings~\cite{Bringmann:2023opz,NANOGrav:2023hvm}, we thus conclude that the impact is minimal --
the main effect being that an additional SMBH contribution generally increases the goodness of fit
somewhat, allowing for a somewhat larger range of phase transition  parameters to be consistent with the data.
The effect is largest for values of $\beta/H\lesssim 10$, where the phase transition  spectrum 
peaks at around $100\,\text{nHz}$ and the PTA data is mostly fitted by the rising part of the spectrum;
the SMBH contribution then helps to accommodate the deviations from a pure power-law shape seen in the data.
In the right panel of figure~\ref{fig:model-independent} we further illustrate this by
showing the best-fit GW spectrum from this combined analysis.
The main point is that the slope of $\Omega_\text{gw}^\text{BH}(f)$ can only vary
between roughly $-1$ and $2$.
Even for amplitudes on the high side compared to
expectations~\cite{NANOGrav:2023hfp}, it is thus difficult to achieve a good fit to the data for more
than the first few frequency bins~\cite{Chen:2025wel}. A phase transition  signal, on the other
hand, can easily accommodate the large slope seen in the data (combined with a clear
prediction of a turn-over towards frequencies somewhat larger than currently accessible).
In summary, the presence of the astrophysical background is very unlikely to affect
conclusions about what is needed for a DSPT to be the dominant cause of the PTA signal, at
least given the current knowledge concerning these backgrounds. In the following, we will
therefore neglect this contribution to the GW signal and only consider the GWs produced
due to the DSPT itself.

Let us finally comment on relevant cosmological constraints for a generic phase transition scenario. First, the GWs 
produced 
due to the transition contribute to the radiation content of the universe, potentially in conflict with tight constraints from 
BBN and CMB that are typically formulated in terms of the number of additional 
effective neutrino degrees of freedom, $\Delta N_\text{eff}\gtrsim 0.22$~\cite{Yeh:2022heq}. 
However, as indicated in figure~\ref{fig:model-independent}, this constraint is irrelevant inside the parameter
region favoured by the PTA signal.
Additional light degrees of freedom in the dark sector, on the other hand, 
are generally a bigger concern. In fact, a phase transition  interpretation 
of the PTA signal can only avoid these constraints
if the dark sector completely thermalises with the visible sector
before the onset of BBN~\cite{Bringmann:2023opz}. Small values of $\xi$ do not help to evade this conclusion either, because the resulting constraint on $\Delta N_\text{eff}$ is driven by the energy density in the dark sector \textit{after} the transition, which is dominated by the required large amount of liberated energy in the phase transition in order to explain the PTA signal. While we have simply assumed thermal equilibrium in this section,
the validity of this assumption will be an important aspect for the discussion of concrete models below.

\section{Dark sector phase transition models}
\label{sec:models}

In the previous section, we outlined the generic requirements on the phenomenological parameters  
$\alpha$, $\beta/H$ and $T_\text{reh}$ that are needed in order to allow for a phase transition  explanation of the PTA signal.  
Fundamentally, these quantities follow from the concrete model that realises such a DSPT. In particular,
they are all related to the effective potential $V_\text{eff}(\phi, T)$ that results from the underlying theory, 
with $\phi$ being the scalar field(s) triggering the PT. The phase-transition strength, to start with, is related to the difference
of the pseudo-trace 
$\bar\theta$ of the energy-momentum tensor in the false and true vacuum, 
at the temperature $T_\text{p}$ of percolation~\cite{Giese:2020znk, Giese:2020rtr},
\begin{align}
    \alpha = \frac{\bar{\theta}_\text{false}(T_\text{p}) - \bar{\theta}_\text{true}(T_\text{p})}{\rho_\text{rad}(T_\text{p}) } \, .
    \label{eq:alpha}
\end{align}
Here, $\rho_\text{rad}$ is the total energy density in radiation (in the SM and dark sector bath) in the symmetric phase,
at the time of percolation, and $\bar{\theta}\approx \big(1-\frac14 T\partial_T\big) V_\text{eff}\big(\!\left<\phi\right>, T\big)$ for $c_\text{s}\approx 1/\sqrt{3}$.
For transitions dominantly triggered by thermal (rather than quantum) fluctuations, all other quantities -- including $T_\text{p}$ itself --
can be derived from the Euclidean bounce action
\begin{align}
 S_3(T) = 4 \pi \int r^2 \, \mathrm{d}r \bb{\frac{\ba{\partial_r \bar{\phi}_\text{b}}^2}{2} + V_\text{eff}(\bar{\phi}_\text{b}, T)} , \label{eq:bounceaction}
\end{align}
where $\bar{\phi}_\text{b}$ is the bounce solution of the $O(3)$-symmetric Klein-Gordon equation for $\phi$.
This directly translates to the bubble nucleation rate per unit time and volume, $ \Gamma(T) = A(T) \exp (- S_3/T)$, 
from which it is in principle straight-forward to calculate the false vacuum fraction $P_\text{f}(T)$~\cite{Athron:2023xlk}.
This, in turn, leads to an implicit definition of $T_\text{p}$ as $P_\text{f}(T_\text{p})=0.71$~\cite{Ellis:2018mja}
and allows to directly compute the length scale $R H_*$~\cite{Athron:2023xlk} which we use for the  determination 
of $\beta/H$ as in eq.~\eqref{eq:bH}.  
In light of the subsequent discussion, we note that a very useful estimate is often given by
\begin{align}
  \label{eq:betaH_approx}
  \beta/H  \approx T \left. \td{S_3}{T} \right|_{T=T_\text{p}} \, ,
\end{align}
a relation which in fact has previously been used as an alternative definition of $\beta/H$.

In practice, the computation of the false vacuum fraction $P_\text{f}(T)$ and hence all derived quantities is 
more involved, as the Hubble rate itself depends on $P_\text{f}(T)$. This is especially relevant when the vacuum
energy released in the transition becomes sizeable. We use the numerical code
\texttt{TransitionListener} to perform all of the above steps in a self-consistent manner,
and refer to ref.~\cite{TL-release} for details concerning the implementation.

To illustrate the procedure and to gain some intuition about  
the expected correlation of $\alpha$ and $\beta/H$ in concrete models, we will
in the following subsection  discuss the case of the thin-wall limit in which the bounce action can be
approximated analytically. We will then turn to three concrete DSPT models.

\subsection{Generic correlation between transition strength and speed}
\label{sec:genericcorrelation}
Let us for simplicity consider the following parametric dependence of the potential difference
between the true and false vacuum, with constants $\kappa$ and $T_\text{c}$:
\begin{align}
  \Delta V = \kappa (T^2 - T_\text{c}^2)\,.
  \label{eq:DeltaV}
\end{align}
This description is motivated by the leading thermal corrections to the effective
potential being proportional to $T^2$, and it is designed such that $\Delta V$ vanishes at the critical 
temperature $T_\text{c}$. In a given model, the temperature dependence
is more intricate, but this simple form captures the most important features for our discussion.

In the thin-wall limit, i.e.~for transitions happening close to criticality $T_\text{p} \lesssim T_\text{c}$,
the field profile $\bar{\phi}(r)$ inside a critical bubble is well-approximated by a step function jumping
from the true vacuum to the false one at the bubble radius $R_0$:
\begin{align}
  S_3(T) = -2 \times 4\pi \int_0^\infty \! r^2 \diff r \,  V(\bar{\phi}(r), T) = \frac{8 \pi}{3} \Delta V R_0^3 = 72 \pi \frac{\sigma^3}{(\Delta V)^2}\,.
     \label{eq:S3_virial}
\end{align}
Here, the first equality follows from an identity for the bounce solution that relates the gradient field
energy to the potential term, similar to the virial theorem in classical mechanics (sometimes called Derrick's or Coleman's trick~\cite{Derrick:1964ww, 
Coleman:1985rnk}). In the last step, the critical bubble radius $R_0 = 3 \sigma / \Delta V$ was used to express the bounce action in terms of the
wall tension $\sigma$. 
The latter characterises the energy cost of forming the bubble wall
due to the height of the potential barrier separating the two minima.
In accordance with the thin-wall limit, we will treat $\sigma$ as being temperature-independent. This is justified
since the dependence of the potential barrier on temperature is
typically much milder than that of $\Delta V$.

Hence, the dominant temperature dependence of the bounce action in our model stems from eq.~\eqref{eq:DeltaV},
and the transition speed approximation from eq.~\eqref{eq:betaH_approx} evaluates to
\begin{align}
  \beta/H \approx \frac{S_3(T_\text{p})}{T_\text{p}}
  \frac{T_\text{c}^2 - 5T_\text{p}^2}{T_\text{p}^2-T_\text{c}^2} \, .
  \label{eq:betaH_T_dependence}
\end{align}
We observe that $\beta/H$ vanishes for a transition happening at $T_\text{p} \to T_\text{c}/\sqrt{5}$ and
increases until it diverges for $T_\text{p} \to T_\text{c}$. The first limit corresponds to metastability,
where the bounce action develops a minimum and the phase transition is very slow.

To estimate the phase transition  strength parameter $\alpha$, we employ eq.~\eqref{eq:alpha} for a radiation energy density
$\rho_\text{rad} = \zeta \, T_\text{p}^4$, where $\zeta \equiv \pi^2 g_*/30$ is proportional to  the 
relativistic degrees of freedom $g_*$
in the thermal bath at temperature $T_\text{p}$. Assuming for simplicity that $g_*$ is temperature-independent, we find
\begin{align}
  \alpha = \frac{\kappa}{2\zeta}\frac{(2T_c^2-T^2_\text{p})}{T^4_\text{p}} \equiv  \alpha_\text{c} \left( 2 - \frac{T_\text{p}^2}{T_\text{c}^2}\right) \frac{T_\text{c}^4}{T^4_\text{p}}\, .
  \label{eq:alpha_T_dependence}
\end{align}
We hence conclude that lower transition temperatures correspond to stronger
transitions: At the critical temperature one finds the value 
$\alpha_\text{c} = \kappa/(2 \zeta T_\text{c}^2)$; close to metastability, $T_\text{p} \to T_\text{c}/\sqrt{5}$,
the maximum strength is reached at $\alpha_\text{m} = 45 \,  \alpha_\text{c}$. 

These relations allow us to predict a strong correlation 
between $\alpha$ and $\beta/H$, which can be made
explicit by eliminating $T_\text{p}$ in eq.~\eqref{eq:betaH_T_dependence},
using eq.~\eqref{eq:alpha_T_dependence}, to obtain
\begin{align}
\frac{\beta}{H} = \frac{S_3(T_\text{p})}{T_\text{p}}
\frac{\alpha - 7 \alpha_\text{c} - 2 \sqrt{\alpha_\text{c} (8 \alpha + \alpha_\text{c})}}{\alpha_\text{c} - \alpha} \, .
\end{align}
The prefactor $S_3(T_\text{p})/T_\text{p}$
is of order $\mathcal{O}(200)$ for transitions happening
at the MeV scale~\cite{Caprini:2019egz}; the precise value depends on the details of the model like particle content and wall velocity, but plays no role in our estimate. 
The remaining
expression encodes the correlation between $\alpha$ and $\beta/H$: 
It has a pole at $\alpha = \alpha_\text{c}$ and decreases monotonically to zero
at $\alpha = 45 \, \alpha_\text{c}$, corresponding to metastability.
\begin{figure}
	\centering
  \includegraphics[width=0.5\linewidth]{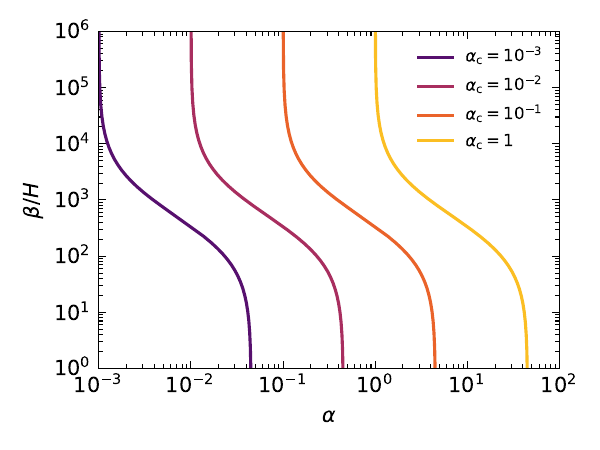}
	\caption{Correlation between transition strength $\alpha$ and speed $\beta/H$
    in the thin-wall approximation for $S_3(T_\text{p})/T_\text{p} = 200$ and different choices of $\alpha_\text{c}$
    as introduced in eq.~\eqref{eq:alpha_T_dependence}.
     See text for details.}
	\label{fig:alphabetaH_analytical}
\end{figure}
In figure~\ref{fig:alphabetaH_analytical} we show this 
correlation between $\alpha$ and $\beta/H$
for different choices of $\alpha_\text{c}$, demonstrating that the limit
of slow transitions corresponds to strong transitions.
However, the figure also clearly illustrates that the
generic prediction -- i.e.~not very close to metastability -- are rather fast transitions with
$\beta/H \gg 100$.
In other words, the {\it generic} expectation from a DSPT is in tension with the requirement to fit the PTA data that we have identified in the previous section,
namely slow and at the same time strong transitions.
This is in line with the findings of the following model-specific
case studies in sections~\ref{sec:u1dark}--\ref{sec:conformal}, which
illustrate the general difficulties of achieving
such slow transitions in concrete models -- and possible ways out in near-conformal settings.

Let us conclude by stressing the limitations of the
above analytical estimates based on our simplistic model:
Most importantly, we relied
on the approximate form of the potential difference in 
eq.~\eqref{eq:DeltaV} and the thin-wall limit of the
bounce action computation. Both assumptions break
down for transitions happening far away from criticality,
rendering the above estimates not applicable to a real model
for $\alpha \gg \alpha_\text{c}$. As derived in
section~\ref{sec:model-independent}, it is however precisely
this regime of thick walls that is required in order to
explain the PTA data. 
Our analytical estimates thus
need to be replaced by numerical computations of the bounce action
and the resulting phase transition parameters. 
In the following subsections, we will demonstrate this explicitly 
for three concrete and representative DSPT models.

\subsection{Abelian dark sector}
\label{sec:u1dark}

As our first example we consider a dark sector with a $U(1)^{\prime}$ gauge symmetry, with associated gauge boson $A^{\prime}_{\mu}$,
and a dark Higgs field $\Phi$ charged under this symmetry:
\begin{align}\label{eq:lagrangian-u1p}
   \mathcal{L} = |D_{\mu}\Phi |^2  - \frac{1}{4} F^{\prime}_{\mu\nu}F^{\prime\mu\nu}  - V(\Phi)\,,
\end{align}
with $F^{\prime}_{\mu\nu}= \partial_\mu  A_\nu^\prime-\partial_\nu  A_\mu^\prime$ and $D_\mu = \partial_\mu + i g A_\mu^\prime$. 
The tree-level scalar potential allowing for a minimum at $\Phi \neq 0$ is
\begin{align}\label{eq:Vu1p}
  V(\Phi) = -\mu^2 \Phi^{*}\Phi+ \lambda (\Phi^{*}\Phi)^2 \,.
\end{align}
To study the phase transition  dynamics in this model, we compute the finite-temperature
effective potential at 1-loop order in perturbation theory. While this is a
standard approach in the literature and sufficient for our purposes,
we note that higher-order and non-perturbative improvements
(e.g., dimensionally reduced effective theories) can reduce gauge and
renormalisation-scale dependence and improve the convergence of the
expansion~\cite{Kajantie:1995dw, Croon:2020cgk}. Since
our focus is on the qualitative model behaviour and broad parameter trends
relevant for the GW signal, we do not expect these refinements to affect
our main conclusions, see also the comparison in ref.~\cite{Lewicki:2024xan}.
The effective potential thus takes the form
\begin{equation}
 V_{\rm eff}(\phi_\text{b}, T) = V_0(\phi_\text{b}) + V_{\rm CW}(\phi_\text{b}) + V_{\rm ct}(\phi_\text{b}) + V_T(\phi_\text{b}, T) + V_{\rm daisy}(\phi_\text{b}, T)\,,\label{eq:effectivepotential}
\end{equation}
based on expanding 
$\Phi = (\phi_\text{b} + \phi + \mathrm{i} \varphi)/\sqrt{2}$ around a (homogeneous and static) 
background field value $\phi_\text{b}$. Here the first term  is the tree-level contribution, 
$V_0(\phi_\text{b})=V(\Phi=\phi_\text{b}/\sqrt{2})$, and the second the usual 1-loop Coleman-Weinberg (CW)
potential~\cite{Coleman:1973jx}, 
\begin{align}\label{eq:coleman-weinberg-potential}
  V_{\mathrm{CW}}(\phi_\text{b}) &= \sum_{a\in \bc{\phi, \varphi, A^\prime}} g_{a} \frac{m_a^4(\phi_\text{b})}{64 \pi^2} \left[ \log \frac{m_{a}^2(\phi_\text{b})}{\Lambda^{2}} - C_{a} \right].
\end{align}
In this expression, we take the renormalisation scale to be the zero-temperature vev, $\Lambda=v\equiv \langle\phi_\text{b}+\phi\rangle_{T=0}=
 \langle \phi_{\rm b} \rangle_{T = 0}$, and we
chose on-shell-like renormalisation conditions~\cite{Basler:2018cwe} enforcing that both vev and background field-dependent masses are given by
their classical tree-level expressions, i.e.~$v=\mu / \sqrt{\lambda}$ and
\begin{align}\label{eq:masses}
  m^2_{\phi}(\phi_\text{b}) = -\mu^2+3\lambda\phi_\text{b}^2\,, && m^2_{\varphi}(\phi_\text{b}) =-\mu^2 + \lambda\phi_\text{b}^2\,,  && m_{A'}^2(\phi_\text{b}) = g^2\phi_\text{b}^2\,.
\end{align}
This fixes the counterterm contribution $V_{\rm ct}$ = $-\delta \mu^2 \phi_\text{b}^2/2 + \delta \lambda \phi_\text{b}^4/4$ in eq.~\eqref{eq:effectivepotential}.
The renormalisation  scale $\Lambda=v$ is chosen such that the logarithms in 
eq.~\eqref{eq:coleman-weinberg-potential} are parametrically small
for field values around the vev, which is the regime relevant for the phase transition  dynamics. Note that this choice would not be 
appropriate for studying
the conformal limit $\mu \to 0$, which we will discuss separately in section~\ref{sec:conformal}.
The internal degrees of freedom are given by $g_{\phi,\varphi} = 1$, $g_{A^\prime} = 3$ and, since 
we employ the $\overline{\text{MS}}$ scheme,  the
renormalisation constants are $C_{\phi, \varphi} = \frac{3}{2}$ and $C_{A^\prime} = \frac{5}{6}$.

The finite-temperature contribution takes the form
\begin{equation}
  \label{eq:VT}
  V_T(\phi_\text{b}) = \frac{T^4}{2\pi^2} \sum_a g_a J_{\text{b}}\left(\frac{m_a^2(\phi_\text{b})}{T^2}\right)\,,
\end{equation}
with the bosonic thermal function $J_{\text{b}}$ defined in ref.~\cite{Quiros:1999jp}. Infrared divergences 
in this latter contribution are taken care of by ``daisy resummation''~\cite{Carrington:1991hz}, for which 
the Arnold-Espinosa prescription~\cite{Arnold:1992rz} leads to
\begin{equation}
  V_{\rm daisy}(\phi_\text{b}) = -\frac{T}{12\pi}\sum_a g_a \left[(m_a^2(\phi_\text{b}) + \Pi_a(T))^{3/2} - (m_a^2(\phi_\text{b}))^{3/2}\right]\,.
\end{equation}
In this last expression the transverse components of $A'$ are {\it not} included in the
final sum over $a$, and the one-loop thermal masses read 
\begin{align}
  \Pi_\phi =  \Pi_\varphi = \left(\frac{\lambda}{3} + \frac{g^2}{4}\right)T^2\,, &&
   \Pi_{A^\prime_L} = \frac{1}{3} g^2T^2 \,.
\end{align}

For sufficiently large couplings $\tilde g \equiv g/\lambda^{1/4}\gtrsim 1$, this model features a phase transition  that
turns from cross-over to first order; increasing $\tilde g$ further leads to stronger and stronger
supercooling until, for $\tilde g\gtrsim 3$, a phase transition  is prevented by false vacuum trapping~\cite{Bringmann:2023iuz}.
Adding (fermionic) matter
fields charged under the $U(1)'$ implies that slightly larger gauge couplings $g$ are needed
for small values of $\lambda$, also to avoid an unstable potential, but otherwise does not change the above
picture qualitatively (see ref.~\cite{Bringmann:2023iuz} for a more detailed discussion).

\begin{figure}[t]
	\centering
    \begin{minipage}{.4\textwidth}
      \begin{tabular}{llc}
        \toprule
        \bf Parameter &  \bf Range & \bf Prior \\
        \midrule
          $\tilde g\equiv g\, \lambda^{-\frac14}$ &  $[0.8,3.2]$ & lin \\
        $\lambda$ & $[10^{-5},1]$ & log \\
        $v \, / \, \text{MeV}$ & $[1,1000]$ & log \\
        \bottomrule
      \end{tabular}
    \end{minipage}~
    \begin{minipage}{.6\textwidth}
      \centering
      \includegraphics[width=\linewidth]{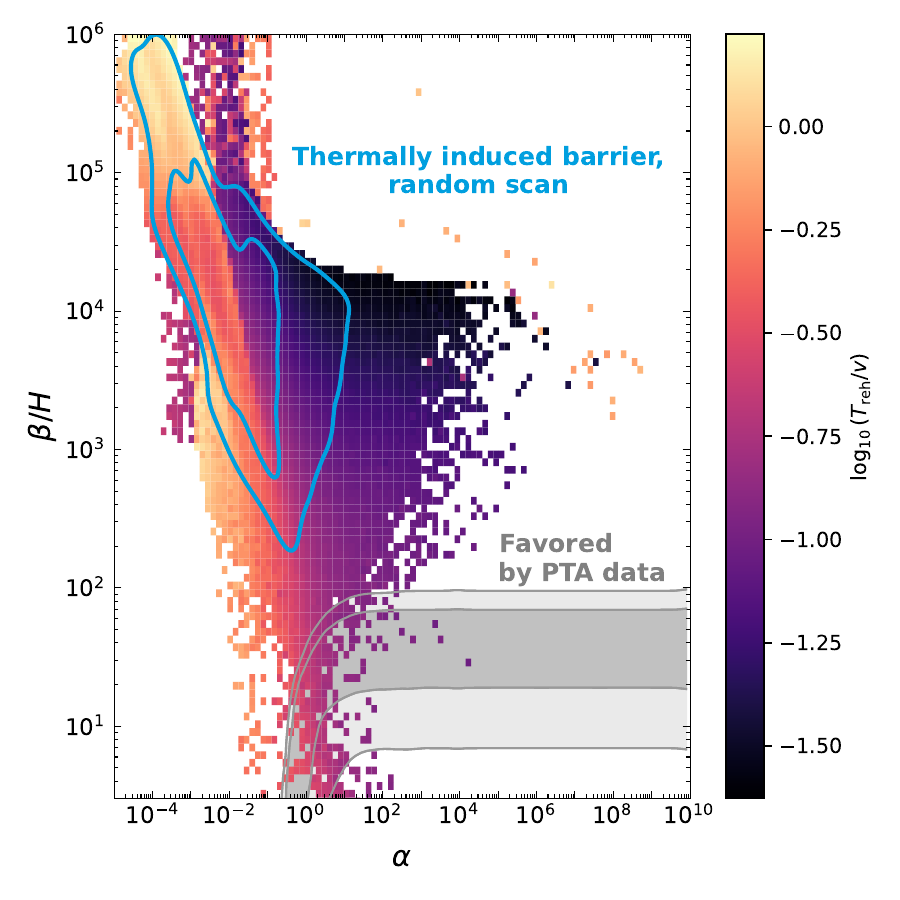}
    \end{minipage}

	\caption{{\it Left panel.} Prior choices for the Abelian dark sector model, 
         targeted at a FOPT at the MeV scale.
	 {\it Right panel.}  The blue lines show the values of $\alpha$ and $\beta/H$ for the produced GW signal that correspond 
	 -- at the $1\sigma$ and $2\sigma$ level -- to the model parameter ranges in the left panel.  For each point in these
	 contours, the colour code indicates the ratio of mean reheating temperature $T_\text{reh}$ and vev $v$. 
	 For comparison, the grey-shaded
	 area shows the result of the model-independent fit to the PTA data, as displayed in 
	 figure~\ref{fig:model-independent}.
	 }
	\label{fig:abelian}
\end{figure}

The above discussion motivates the range of parameters that we are interested in to describe a FOPT at the MeV scale,
and which we summarise in the left panel of figure~\ref{fig:abelian}. In particular, we choose to scan directly over 
$v$ rather than $\mu$, as the only dimensionful parameter in the model, 
given that $v$ is more directly connected to the scale at which the 
transition takes place (recalling that $T_\text{p}\sim v$). The range of $\lambda$, on the other hand, has only limited impact 
on the phenomenology we are interested in here.
In the right panel, we show the resulting range of $\alpha$ and $\beta/H$, along with the mean (log) reheating
temperature for each point in this plane. As expected, the ratio $v/T_\text{reh}$ grows with $\alpha$, as the phase transition  becomes
stronger (and more supercooled). 
For comparison, we also show the results of the model-independent PTA fit that we discussed in 
Section~\ref{sec:model-independent}. Clearly, the generic prediction of the model lies outside
the region in the $\alpha - \beta/H$ plane favoured by the PTA data, with overlap occurring only
at (at least) the $3\sigma$ level. In other words, and as 
discussed in more detail in Section~\ref{sec:results}, significant tuning of the model parameters
is required to explain the data with a phase transition  in an Abelian dark sector.

\begin{figure}[t]
	\centering
	\vspace*{4cm}
  \includegraphics[width=\linewidth]{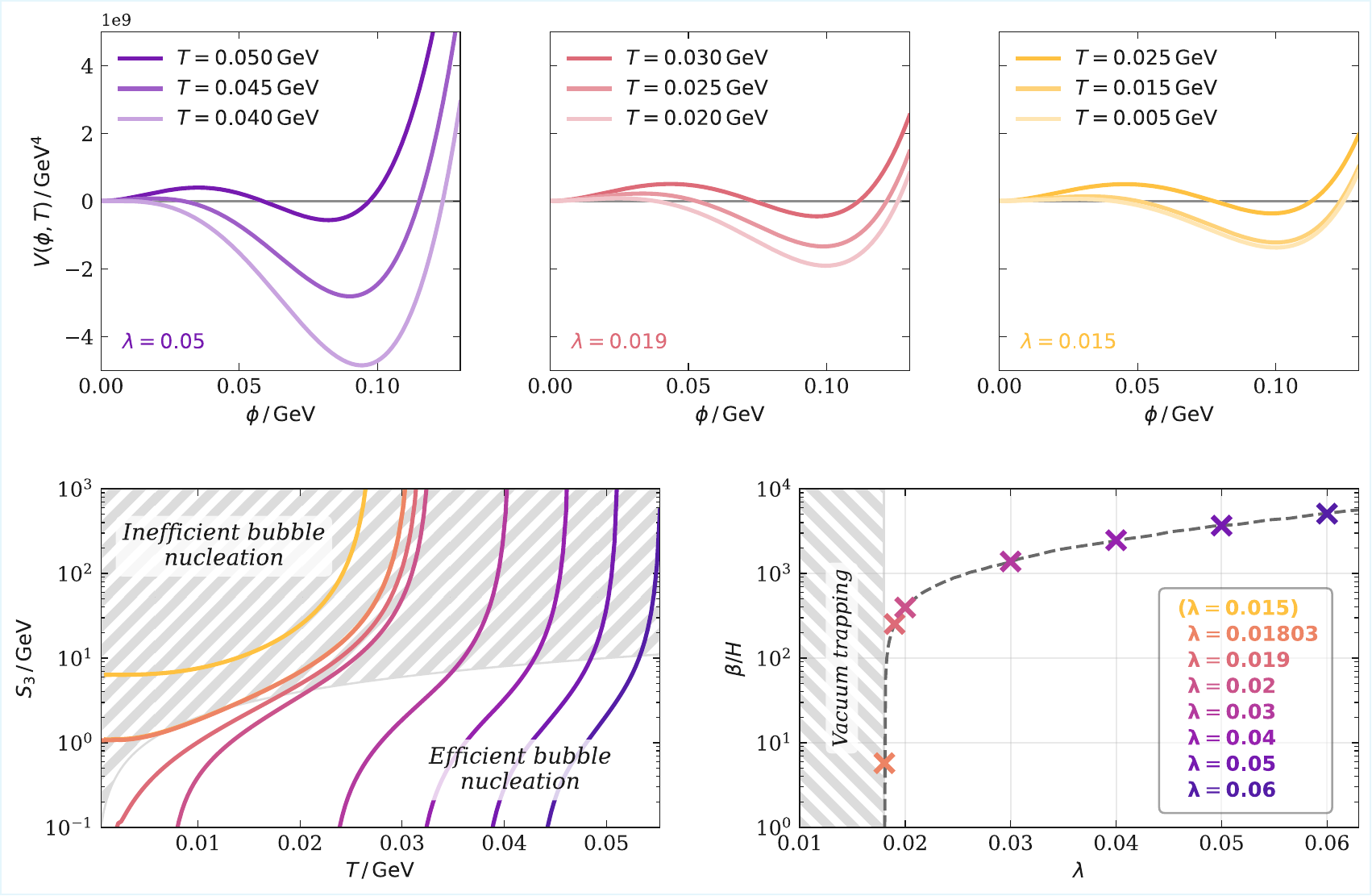}
	\caption{\textit{Upper row:} Effective potential of the 
	Abelian dark sector model at different temperatures,  choosing for illustration model parameters 
	$g=1$, $v = 100 \, \text{MeV}$ and three 
	different values of the quartic coupling $\lambda$ as indicated.
        \textit{Lower row, left:} Corresponding tunnelling action $S_3$ as a function of temperature. 
        \textit{Lower row, right:} Resulting inverse transition timescale $\beta/H$ as a function of the quartic coupling $\lambda$.}
	\label{fig:S3_U1}
\end{figure}

In figure~\ref{fig:S3_U1} we illustrate why it is so difficult to identify model parameters that result in GW parameters
in the desired regime. The three panels in the top row show the effective potential for three different values
of the quartic coupling $\lambda$ (keeping $g$ and $v$ fixed), for a selection of temperatures $T$.\footnote{%
Here we define the false vacuum to correspond to $V=0$, i.e.~we subtract $V_{\rm eff}(\phi=0, T)$ 
from eq.~\eqref{eq:effectivepotential}. 
} 
In general, the potential barrier decreases with temperature, implying a correspondingly smaller tunnelling action
and hence more efficient bubble nucleation. A smaller value of $\lambda$, as visible when comparing the top panels 
from left to right, leads to a less rapid decrease of the potential barrier with $T$ (because the coefficient of the quartic
term in the potential only depends logarithmically on the temperature). Bubble nucleation thus becomes less and less 
rapidly efficient for smaller temperatures
and, for very small values of $\lambda$, nucleation will in fact become completely inefficient even in the $T\to0$ limit. 
This is exactly the behaviour that is illustrated in the bottom left panel: for small values of $\lambda$, 
the Euclidean action always remains at $S_3/T > 200$ (indicated as hatched area).  Since the transition rate is
basically given by the slope of $S_3/T$, cf.~eq.~\eqref{eq:betaH_approx}, values of $\beta/H$ that are sufficiently 
small to explain the PTA data can thus only be achieved for a tiny range of $\lambda$. We show this explicitly
in the bottom right panel: for values of $\lambda$ a bit larger than $\sim0.018$, the phase transition  proceeds too fast -- 
while for values only slightly smaller, the phase transition  does not successfully complete at all. We note that in our 
parameterisation the necessarily strong tuning that is associated with this observation appears in the quantity 
$\tilde g \equiv g/\lambda^{1/4}$, rather than $\lambda$, see also the discussion in ref.~\cite{Costa:2025csj}.

So far, we have only discussed the GWs produced by the phase transition, 
tacitly assuming that the dark sector 
temperature equals that of the visible sector. As mentioned in section~\ref{sec:model-independent},
the dark sector must decay sufficiently quickly afterwards to 
avoid the stringent BBN constraints on $\Delta N_{\rm eff}$.
For the Abelian model, the easiest way to realise this is by assuming the usual portal couplings to be non-vanishing. 
Specifically, the dark photon (dark Higgs) could decay 
via kinetic mixing (Higgs mixing) consistent with complementary constraints.

Let us start our discussion with a closer inspection of the produced \textbf{dark photons}. We will find later that the preferred 
dark photon masses in this scenario are in the range  $ 10 \, \text{MeV} \lesssim m_{A'} \lesssim 2 \, \text{GeV}$
(see figure~\ref{fig:corner_thermal_barrier} in appendix \ref{app:triangles}). 
This implies that in all cases under consideration electrically charged final states are kinematically available. 
In order to be compatible with BBN a lifetime $\tau \lesssim 1$\,s is hence required (see ref.~\cite{Depta:2020zbh} for 
more detailed bounds also taking into account the abundance of the relic). 
For $m_{A'} \lesssim 200\,\mathrm{MeV}$, the main decay channel will be $A' \to e^+ e^-$ and $\tau \lesssim 1$\,s 
corresponds to a 
kinetic mixing $\epsilon \gtrsim 10^{-9} - 10^{-10}$. This is very close to or may even be in conflict with existing 
supernova limits~\cite{Chang:2016ntp}, in particular for masses where cooling in the gain layer is 
relevant~\cite{Caputo:2025aac}.
There is however an open (slightly mass-dependent) window for $10^{-6} \lesssim \epsilon \lesssim 10^{-4}$ between 
collider and beam dump limits for $m_{A'} \gtrsim$ a few MeV. While a viable possibility today, this region may be probed 
in the near future in particular by Belle II~\cite{Ferber:2022ewf}, NA64$\mu$~\cite{Gninenko:2653581} and 
LHCb~\cite{LHCb:2022ine}, but also by FASER2~\cite{Anchordoqui:2021ghd}, DarkQUEST \cite{Apyan:2022tsd} and 
SHiP~\cite{Ahdida:2023okr}. For even heavier dark photon masses,  $m_{A'} \gtrsim 200\,\mathrm{MeV}$, the bounds 
become much weaker, as additional
decay channels open up (making the decay quicker for fixed kinetic mixing $\epsilon$) 
while at the same time the dark photons become too heavy to be produced in supernovae.

Turning to the  \textbf{dark Higgs}, we observe that the preferred masses are somewhat lighter, 
$2  \, \text{MeV} \lesssim m_\phi \lesssim 1 \, \text{GeV}$ (see again figure~\ref{fig:corner_thermal_barrier} in 
appendix \ref{app:triangles}). 
We also find that for all parameter points of interest $m_\phi < 2 m_{A'}$,
so that the dark Higgs cannot decay into two dark photons.
To allow for a consistent phenomenology we therefore need to consider other possibilities, the simplest one 
corresponding to an induced mixing with the SM Higgs via the usual portal coupling
\begin{align}\label{eq:higgs-mixing}
  \mathcal{L}_{\mathrm{portal}} = \lambda_{H\Phi} |H|^2 \Phi^*\Phi\,.
\end{align}
After electroweak symmetry breaking, such a term
would induce a non-vanishing mixing angle $\sin\theta \simeq \lambda_{H\Phi}\,\frac{v\,v_{\mathrm{EW}}}{m_h^2}$
(for $m_\phi \ll m_h$), which would allow for $\phi$ decays into SM final states. 
The Higgs portal coupling would also induce an additional effective mass term for $\phi$ below the electroweak scale, 
which however can be absorbed into a redefinition of the tree-level parameters without changing the dynamics of the 
electroweak and dark sector phase transitions.
Two separate phase transitions $(0,0) \to (0,v_\text{EW}) \to (v, v_\text{EW})$ can be ensured 
by requiring that the curvature in the $\phi$ direction of the effective potential 
remains positive for 
temperatures around the electroweak scale, 
i.e.~$\left. \partial_\phi^2 V_{\rm eff}(h,\phi, T \simeq v_\text{EW})\right|_{\phi = 0} > 0$. This leads to an upper bound on the portal coupling $\lambda_{H\Phi} \ll \lambda$, or equivalently
\begin{align} \label{eq:portal-bound}
  \sin \theta \ll \ba{\frac{m_\phi}{m_h}}^2 \ba{\frac{v_\text{EW}}{v}} \simeq 10^{-4} \ba{\frac{m_\phi}{10 \, \text{MeV}}}^2 \ba{\frac{10 \, \text{MeV}}{v}} \, ,
\end{align}
where $m_h = 125$ GeV and $v_\text{EW} = 246$ GeV are the SM Higgs mass and vev, respectively. The analogous 
condition for the dark sector phase transition to not be affected by electroweak scale physics, 
$\left. \partial_h^2 V(h,\phi, T \simeq v)\right|_{h = v_\text{EW}} > 0$, is then
automatically satisfied for $v \ll v_\text{EW}$. For portal couplings satisfying the bound~\eqref{eq:portal-bound}, the 
tree-level mass induced by electroweak symmetry breaking can thus be compensated by a redefinition of the tree-level 
parameters without affecting the phase transition dynamics. This a posteriori justifies our neglect of the portal coupling in 
the analysis of the phase transition dynamics.

We find that for the given mass range $\phi$ decays via Higgs mixing can be efficient enough 
while respecting current constraints. Specifically, for  
$2  \, \mathrm{MeV} \lesssim m_{\phi} \lesssim 2 m_\mu$ the allowed range of $\sin\theta$ is between 
$10^{-5} \le \sin\theta \le 10^{-4}$, where the lower bound comes from the requirement that the dark Higgs decays 
sufficiently early~\cite{Fradette:2018hhl} while the upper limit is due to collider and beam dump 
experiments~\cite{NA62:2021zjw}. Using eq.~\eqref{eq:portal-bound} we arrive at the conclusion that this range is 
consistent with the requirement of two separate phase transitions for almost all of the points shown in 
figure~\ref{fig:abelian}. Once the dark Higgs is sufficiently heavy to decay 
into muons,  cosmological constraints on the mixing angle are significantly 
relaxed. In particular for masses between the muon and pion threshold values as small as $\theta \simeq 10^{-9}$ are 
possible. 
Overall we therefore find that the minimal addition of kinetic and Higgs mixing is sufficient to make the scenario 
cosmologically viable without impacting the PT dynamics.

Let us finally remark that the above requirements of a sufficiently fast decay of the DS degrees of freedom 
automatically guarantees that the dark and visible sector are kept in thermal equilibrium throughout the transition, 
thus justifying our treatment of assuming $T_\text{DS}=T$ throughout the analysis (for a more detailed discussion see,  
ref.~\cite{Bringmann:2023iuz}).

\subsection{Flip-flop}
\label{sec:flipflop}

As our second example, we introduce a two dark singlet model to realize a two-step
phase transition. It consists of two real scalars $\phi_{1}$ and $\phi_{2}$ as well as a Majorana fermion
$\psi=\psi^{c}$, which only couples to $\phi_{1}$:
\begin{align}\label{eq:flipflop-lagrangian}
  \mathcal{L} = \frac{1}{2}\left(\partial_{\mu}\phi_{1}\right)\left(\partial^{\mu}\phi_{1}\right) + \frac{1}{2}\left(\partial_{\mu}\phi_{2}\right)\left(\partial^{\mu}\phi_{2}\right) +
   \frac{1}{2} \mathrm{i} \bar{\psi}\slashed{\partial} \psi - \frac{y}{2} \phi_{1}\bar{\psi} \psi
  - V(\phi_{1}, \phi_{2}) \, .
\end{align}
As indicated, we thus assume the bare
Majorana masses to be sufficiently small to not affect the phenomenology of the phase
transition discussed below. We parametrise the tree-level potential of the two scalars
in the suggestive form
\begin{align}
  \label{eq:flipflop-potential}
  V(\phi_1, \phi_2) = \frac{\lambda_0}{4} \left( \phi_1^2 + \gamma^2 \phi_2^2 - v^2\right)^2 -
  \frac{\lambda_1}{2} v^2 \phi_1^2 + \frac{\lambda_{12}}{2} \phi_1^2 \phi_2^2\,.
\end{align}
The potential is constructed such that its minima,
$(\phi_{1},\phi_{2}) = ( \pm \sqrt{1 + \lambda_{1}/\lambda_{0}}v, 0)$ and
$(\phi_{1},\phi_{2}) = (0, \pm v/\gamma)$, are located on the $\phi_{1}$- and
$\phi_{2}$-axis, respectively. For $\lambda_{1} > 0$ the minimum in the $\phi_{1}$ direction is deeper
than the one on the $\phi_{2}$-axis at zero temperature. The coupling $\lambda_{12}$ between the
two scalars creates a tree-level barrier separating the two minima for sufficiently large
$\lambda_{12} > \gamma^2 \lambda_{1}$. In order to avoid the phenomenological relevance of domain walls, we further 
assume some small bias terms, which favour the positive valued minima on both axes, but otherwise
do not play any role in our analysis.

We compute the effective potential analogously to the
previous section, for more details on the renormalisation and thermal masses see
appendix~\ref{app:flipflop}.
The quantum and thermal corrections to the tree-level potential do not push the minima off
the two field axes; they merely determine which minimum is the global one at a given
temperature, and where on the axes the minima are located. At very high temperatures, the
thermal corrections dominate and there is only a single phase at the origin in field
space. As the temperature decreases, two scenarios are in principle possible: Either the minimum in the
$\phi_{1}$ direction appears first, and becomes the global minimum, or the minimum in the
$\phi_{2}$ direction does so. We are interested in the latter case, which can be achieved by
the fermion obtaining a sufficiently large mass $m_\psi = y \langle\phi_1\rangle /\sqrt{2}$, depending only on the vev in the
$\phi_{1}$ direction. That way, the thermal corrections of the phase on the
$\phi_{1}$-axis are stronger than those on the $\phi_{2}$-axis. The minimum in the
$\phi_{2}$ direction then develops first; subsequently, the minimum in the $\phi_{1}$ direction changes
from a local to the global minimum. If $\lambda_{12}$ is not too small compared to
$\lambda_0$ and $\lambda_1$, the tree-level barrier between the two minima ensures that the second
transition step can be supercooled. 

\begin{figure}[t]
  \centering
  \vspace*{4cm}
  \includegraphics[width=\linewidth]{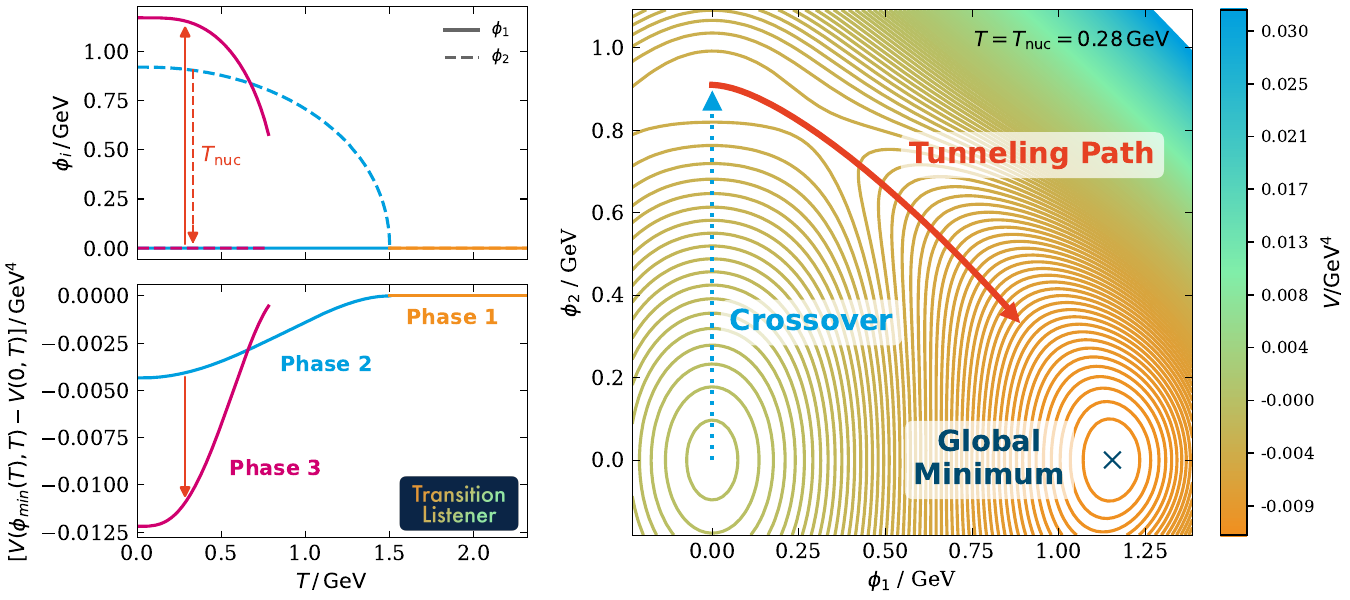}
  \caption{\textit{Left panels}: Field values of $\phi_{1}$ and $\phi_{2}$ as a function of
    temperature (upper panel) and the corresponding potential values (lower panel). At
    high temperatures $T$ the global minimum is at the origin (orange line). At intermediate
    $T$ the global potential minimum smoothly shifts from $(\phi_1,\phi_2)=(0,0)$ to
    $\phi_1 = 0, \phi_2 \neq 0$ with decreasing $T$ (blue line). At low temperatures, a local minimum at
    $\phi_1 \neq 0, \phi_2 = 0$ appears (red line), which eventually becomes the global potential
    minimum. \textit{Right panel}: Effective potential at the nucleation
    temperature $T_{\mathrm{nuc}} = 0.28\,$GeV. The blue dotted arrow shows the evolution of
    the $\phi_{2}$ minimum until $T_{\mathrm{nuc}}$, the red arrow shows the tunnelling path
    from the local minimum to the release point of the field.
    The parameter values for the benchmark point shown are stated in table~\ref{tab:benchmarks}
    in the appendix.
  }
  \label{fig:flipflop_potential}
\end{figure}

The phase structure of such a flip-flop transition is shown for a benchmark point
(table~\ref{tab:benchmarks}) in the left panels of figure~\ref{fig:flipflop_potential}. 
The right panel shows the effective potential at the nucleation temperature, with the
red arrow indicating the tunnelling path. Since the bounce solution lies in the thick wall
regime, the release point of the field (end of the arrow) does not coincide with the
global minimum of the potential.

\begin{figure}[t]
  \centering
  \begin{minipage}{.4\textwidth}
    \begin{tabular}{llc}
\toprule
\bf Parameter &  \bf Range & \bf Prior \\[1pt]
\midrule
$\lambda_0$ &  $[10^{-3},1]$ & log \\[5pt]
$\lambda_1$ & $[10^{-3},1]$ & log \\[5pt]
$\lambda_{12}$ & $[10^{-3},1]$ & log \\[5pt]
$\gamma$ & $[0.5,2]$ & log \\[5pt]
$y$ & $[0.1,\sqrt{4 \pi}]$ & log \\[5pt]
$v \, / \, \text{MeV}$ & $[10,10^4]$ & log \\[2pt]
\bottomrule
\end{tabular}
\end{minipage}~~~
\begin{minipage}{.6\textwidth}
  \vspace*{1cm} \hfill
  \includegraphics[width=\linewidth]{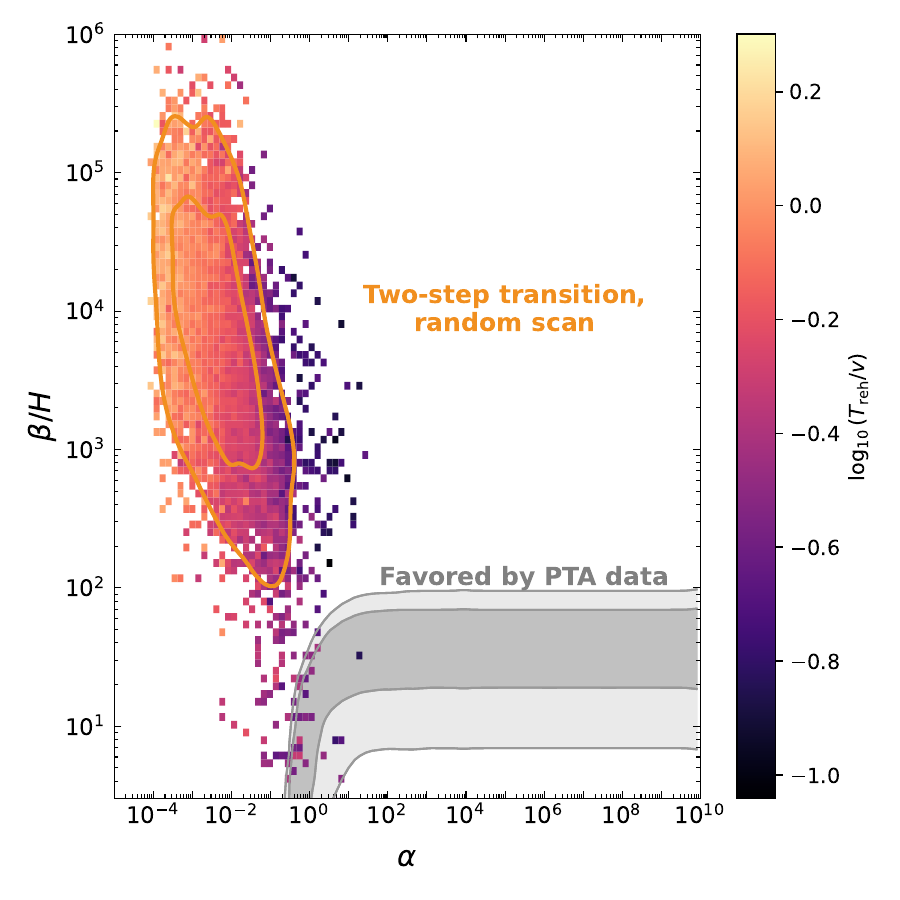}
  
\end{minipage}
\caption{{\it Left panel.} Prior choices for the flip-flop model, targeted at a FOPT at
  the MeV scale. {\it Right panel.} Resulting 1$\sigma$ and 2$\sigma$ contours in the
  $\alpha - \beta/H$ plane of the produced GW signal, with colour coding indicating $T_\text{reh}$.
  For each point in these
	 contours, the colour code shows the ratio of mean reheating temperature $T_\text{reh}$ and vev $v$. 
	 For comparison, the grey-shaded
	 area corresponds to the result of the model-independent fit to the PTA data, as displayed in 
	 figure~\ref{fig:model-independent}.}
\label{fig:flipflop}
\end{figure}

The parameter ranges where we search for two-step transitions are summarised in the left
panel of figure~\ref{fig:flipflop}. Note that not all of the points in this parameter
space will be physically viable or feature a two-step transition. In our scan, we thus explicitly 
exclude points where the zero-temperature vacuum is unstable or where the tunnelling action 
does not drop sufficiently  to allow for a successful transition to the global minimum until today, 
see for reference also the posteriors in figure~\ref{fig:corner_two_step} in appendix~\ref{app:triangles}.
The right side of figure~\ref{fig:flipflop} shows the corresponding range of transition strength $\alpha$ and speed
$\beta/H$,  
with a colour code for the mean (log) reheating temperature at each grid point.
It is clear that only a very
small fraction of the parameter space gives rise to the $\alpha$ and $\beta/H$ values favoured
by the PTA data (grey contour). The minima of the first-order transition are in fact further
separated in field space than in the case of the Abelian model. This increases the bounce
action, suppresses the nucleation rate, and thus leads to smaller $\alpha$ for similar
$\beta/H$ compared to the dark $U(1)'$ model from section~\ref{sec:u1dark}.

Just as we discussed in the case of the $U(1)'$ model, 
the dark sector has to decay sufficiently quickly
to avoid constraints on $\Delta N_{\mathrm{eff}}$ from BBN.
The fermion $\psi$ is typically the heaviest DS state, and hence strongly Boltzmann suppressed at temperatures $T\sim m_{\phi_i}$. In the absence of
gauge bosons, the only renormalisable operators are Higgs portal couplings with the scalars as described around eq.~\eqref{eq:higgs-mixing} above.
For the case of the two scalars considered here, this slightly generalises to two possible couplings, $\lambda_{H\phi_i} |H|^2 \phi_{i}^{2}$, but the phenomenology is still very similar to the case of the Abelian Higgs described above:
After the dark sector transition into the global minimum, the dark singlets can decay via Higgs mixing into 
electrons (and all other charged final states once the corresponding mass thresholds are crossed), consistent with cosmological and complementary constraints.

\subsection{Conformal dark sector}
\label{sec:conformal}

As the simplest example of a classically conformal dark sector, we consider the same
Lagrangian as in eqs.~(\ref{eq:lagrangian-u1p}, \ref{eq:Vu1p}) but without the dimensionful
parameter $\mu$, i.e.
\begin{align}\label{eq:lagrangian-conformal}
   \mathcal{L} = |D_{\mu}\Phi |^2  - \frac{1}{4} F^{\prime}_{\mu\nu}F^{\prime\mu\nu}  - \lambda (\Phi^{*}\Phi)^2 \,.
\end{align}
The conformal symmetry of this model is broken at the one-loop
level~\cite{Coleman:1973jx}, inducing a non-vanishing vev $v\equiv\langle\phi_\text{b}\rangle_{T=0}$. Taking into
consideration the enhanced symmetry, we change the renormalisation condition:
Rather than enforcing the tree-level expressions for masses and vev at $T=0$, as in
section~\ref{sec:u1dark}, we now demand that the mass of the scalar field $\phi_\text{b}$
vanishes at $\phi = 0$, i.e.~$m^2_\phi(\phi_{\rm b}=0)=\partial_{\phi_\text{b}}^2 V_\text{eff}(\phi_\text{b})\big|_{\phi_\text{b}=0}=0$ in the
zero-temperature limit, and define the four-point coupling at some energy scale $\Lambda$ by requiring
$ 6 \lambda=\partial_{\phi_\text{b}}^4 V_\text{eff}(\phi_\text{b})\big|_{\phi_\text{b}=\Lambda}$ at $T\to0$.
We take this energy scale to correspond to the (loop-induced) vev,
$\Lambda=v$, which now differs from the tree-level expression $v=0$. This prescription implies the
relation~\cite{Balan:2025uke}
\begin{equation}
\label{eq:lambda}
\lambda=\frac{11}{48\pi^2}\left(10\lambda^2 + 3g^4 \right),
\end{equation}
which determines $\lambda$ as a function of $g$. Similar to the
$U(1)'$ case discussed in section~\ref{sec:u1dark}, adding fermions does not have a significant impact on the phase 
transition itself; as discussed below, however, it can to some extent help to get rid of the excess
energy density and potentially provide a DM candidate~\cite{Balan:2025uke}.

At high temperatures, the term $T^2 \phi^2$ in the
$V_T$ contribution to the effective potential, cf.~eq.~\eqref{eq:VT}, restores the
$U(1)'$ symmetry. In contrast to the
$U(1)'$ model discussed above, however, 
the barrier now originates from the running of the quartic coupling 
rather than a thermal cubic term that is produced from gauge bosons.
In terms of operators, the running will produce a term of order $\phi^4 \log(\phi^2)$
which together with the quadratic and quartic terms leads to two 
local minima in the potential. Since the effect is logarithmic, 
the scalar field value at the maximum of the barrier is typically much smaller than the
scalar vev in the global minimum. This leads to lower phase 
transition temperatures and hence stronger phase transitions, see also ref.~\cite{Kierkla:2023von}.
For sufficiently large gauge couplings
$g$ the running of $\lambda$ is sizeable and the conformal model will thus always result in a FOPT.

\begin{figure}[t]
	\centering
    \begin{minipage}{.49\textwidth}
   \begin{tabular}{lll}
   \hline
   \hline\\[-4mm]
   \bf Parameter &  \bf Range & \bf Prior \\[1pt]
   \hline\\[-4mm]
    $g$ &  $[0.5,2]$ & linear \\[5pt]
   $v$ & $[1,1000]\,{\rm MeV}$ & logarithmic \\[2pt]
   \hline
   \hline
   \end{tabular}
    \end{minipage}~~~~~
    \begin{minipage}{.49\textwidth}
        \centering
	\vspace*{1cm} \hfill
        \includegraphics[width=\linewidth]{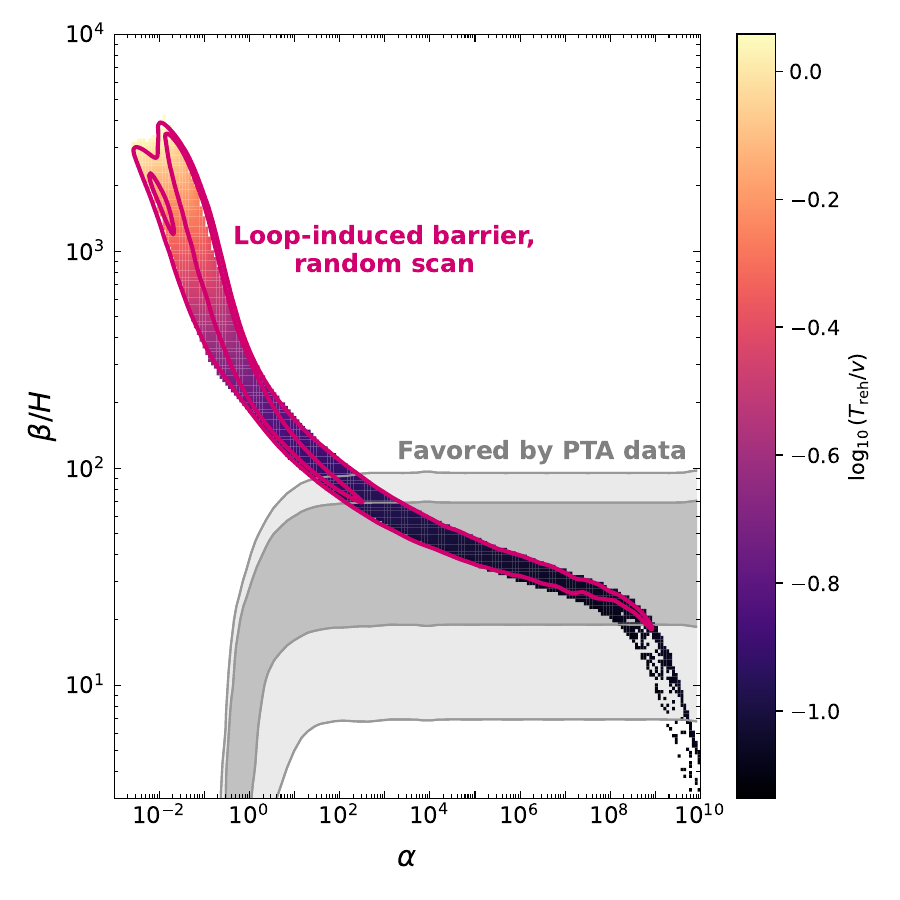}
    \end{minipage}%

	\caption{{\it Left panel.} Prior choices for the conformal dark sector model, 
	at the MeV scale.
	 {\it Right panel.} Resulting 1$\sigma$ and 2$\sigma$ contours in the $\alpha - \beta/H$ plane 
	 of the produced GW signal, with colour coding indicating $T_\text{reh}$. For each point in these
	 contours, the colour code shows the ratio of mean reheating temperature $T_\text{reh}$ and vev $v$. 
	 For comparison, the grey-shaded
	 area corresponds to the result of the model-independent fit to the PTA data, as displayed in 
	 figure~\ref{fig:model-independent}.}
	\label{fig:conformal}
\end{figure}

We summarise the corresponding range of parameters that are potentially interesting for a
FOPT at the MeV scale in the left panel of figure~\ref{fig:conformal}. For values of
$g \gtrsim 0.5$ the model allows for first-order phase transitions; for lower values of $g$,
the transition does not succeed ($P_\text{f} < 0.01$) until today due to vacuum trapping.
The transition becomes weaker and proceeds more quickly as $g$ grows, up to roughly $g\approx2$,
beyond which the quadratic relation~\eqref{eq:lambda} for $\lambda$ admits no real solution and
perturbation theory breaks down. In the right panel of figure~\ref{fig:conformal}
we show the values of $\alpha$ and $\beta/H$ that result from a random scan over these parameter 
ranges, with
the 1$\sigma$ and 2$\sigma$ contours indicated by red lines.  We observe a clear
anti-correlation between $\alpha$ and $\beta/H$, similar to the one expected from
the analytical arguments presented in section~\ref{sec:genericcorrelation}.
A significant fraction of the resulting phase transitions are
strong enough ($10^{2}\lesssim \alpha \lesssim 10^{8}$) and sufficiently slow
($20 \lesssim\beta/H\lesssim 100$) to lie in the region favoured by the PTA data.
As in the previous two examples, the reheating temperature after the phase
transition is typically of order a few hundred MeV and consistent with the expectations that
$T_{\mathrm{reh}} / v$ drops for stronger transitions. 

The decay of the dark sector prior to BBN is again a necessary requirement in order to
avoid stringent constraints on $\Delta N_{\mathrm{eff}}$. In principle, the same decay
channels as discussed in section~\ref{sec:u1dark} could be employed to
achieve this. In contrast to the other cases, however, in the conformal scenario couplings to
the SM cannot easily be realised via a Higgs portal interaction due to the associated induced effective mass below the 
electroweak symmetry breaking scale:
a non-zero portal coupling would  
violate the assumption of near-conformality and thus significantly modify the phase
transition dynamics.
While this may result in a viable phenomenology, we do not consider this possibility 
here and restrict our discussion to the near-conformal framework.
 
We find that some additions to the minimal conformal model are required to achieve a scenario
consistent with cosmological constraints.
The arguably simplest solution would be to add a second Abelian gauge group $U(1)_2^\prime$ that $\phi$ is charged 
under and that mixes kinetically with $U(1)_\text{EM}$, but with a significantly smaller gauge coupling 
$g_2 \ll g$. We checked that for a large range of gauge couplings $g_2$ this will not significantly change the PT 
dynamics, while 
allowing the dark photons $A_2'$ to be still sufficiently heavy for decays $A_2' \to e^+ e^-$ via kinetic 
mixing. The DS can then efficiently transfer its energy density to the SM via dark Higgs decays $\phi \to A_2' A_2'$ 
followed by $A_2' \to e^+ e^-$, thereby guaranteeing a consistent cosmological setup.
Other possibilities that allow the dark sector to decay sufficiently quickly before BBN have been 
discussed in refs.~\cite{Madge:2023dxc,Balan:2025uke}.

\section{Confronting models with data}
\label{sec:results}

\begin{figure}[t]
	\centering
	\includegraphics[width=\linewidth]{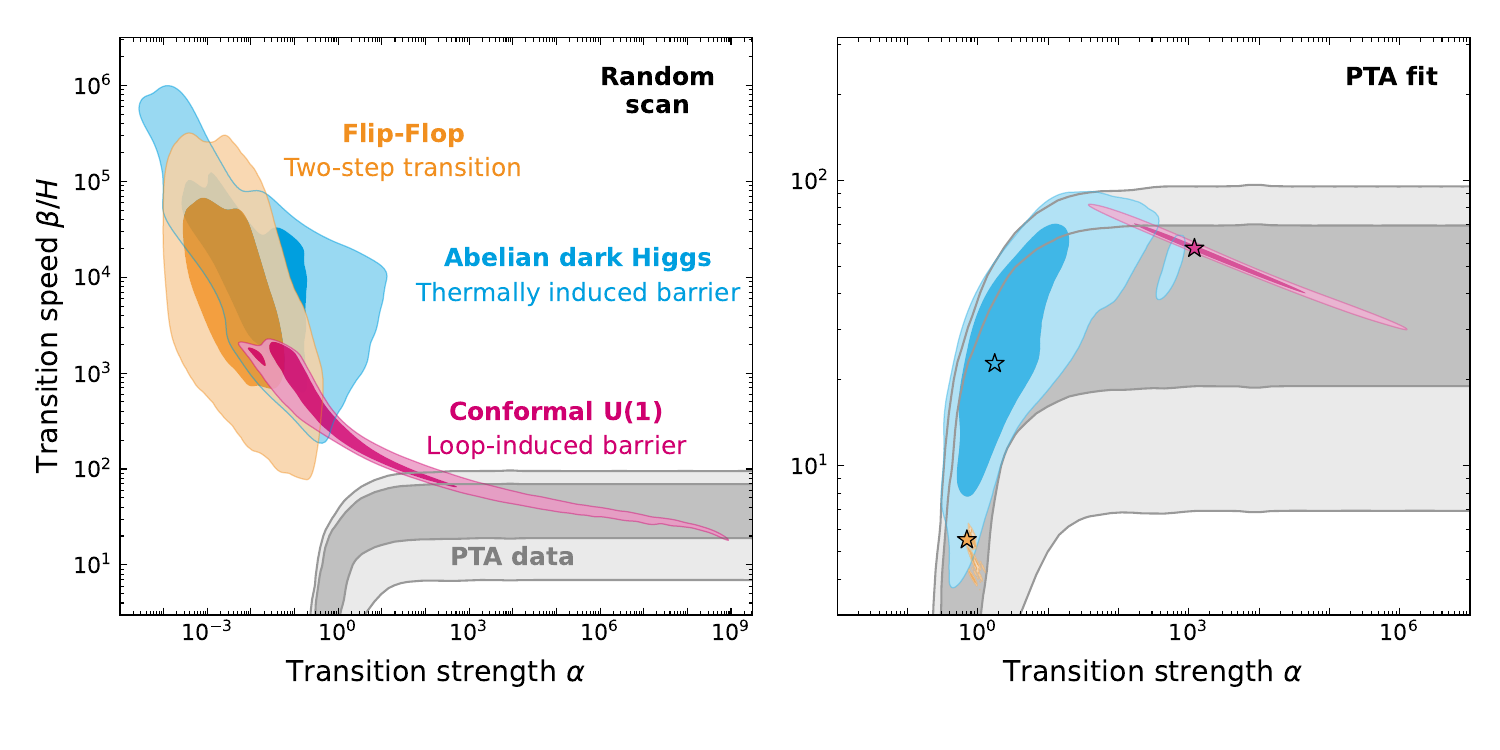}
	\caption{\textit{Left panel:} Generic predictions of our three sample models in the $\alpha - \beta/H$ plane 
	(coloured areas), as displayed earlier in figures~\ref{fig:abelian}, \ref{fig:flipflop} and \ref{fig:conformal}. For
	comparison, the grey-shaded area shows the results for the model-independent fit to the PTA data, 
	cf.~figure~\ref{fig:model-independent}. \textit{Right panel:} Preferred regions for the same models 
	after sampling over the PTA likelihood, with stars indicating the respective best-fit points
	(cf.~tables \ref{tab:benchmarks} and \ref{tab:benchmarks-observables} in appendix~\ref{app:triangles}).
        Note that the contours in the right panel are only indicative for the Abelian Higgs and the flip-flop model, 
        see also footnote~\ref{foot:sampling}, and that the different models require very different
        amounts of tuning to explain the PTA data, see the discussion around table~\ref{tab:tuning}.}	
	\label{fig:summaryplot}
\end{figure}

In the previous section, we have discussed at length the three sample models that we considered as possible
representatives of the type of broader model classes that may explain the GW signal observed by PTAs. The generic 
GW predictions of these models, in particular,
were shown in figures~\ref{fig:abelian}, \ref{fig:flipflop} and \ref{fig:conformal}. For ease of comparison, we combine 
them in the left panel of figure~\ref{fig:summaryplot} in the $\alpha - \beta/H$ plane, along with the results of the 
model-independent analysis of the PTA data presented in Section~\ref{sec:model-independent}. Already a quick 
glance at this figure indicates that it should indeed be straight-forward to explain the PTA signal in terms of a phase 
transition  that 
results from the breaking of a conformal symmetry, with a loop-induced barrier, 
while it will be significantly more challenging to achieve the same for the other options, based on a thermally induced
barrier and a two-step transition, respectively.

In order to test this expectation, we next sample over the full PTA likelihood described
in Section~\ref{sec:model-independent}. Due to the \texttt{PTArcade} implementation of the
PTA likelihood being unphysical below a certain threshold, we extend this likelihood to a
function that smoothly approaches the \texttt{PTArcade} likelihood in the region of
interest, while driving the \texttt{ultranest}~\cite{Buchner:2021cql} sampler away from
uninteresting likelihood plateau regions. 
In view of the numerically highly challenging sampling procedure,
we further adopt our  strategy to identify parameters with a good fit to the PTA
data. For the Abelian dark Higgs model, this means that we only scan over
the restricted range $10^{-3} \leq \lambda \leq 1$. In the case of the two-step model,
we fix $y = \gamma = 1$ and sample only over the remaining four parameters;
for this setup, we start with an initial \texttt{ultranest} scan and then use the 
the 100 (unconverged) samples with the highest likelihood to initialise an MCMC sampler.

In all cases, we further exclude parameter points  that do not lead to a
successful phase transition, i.e.~where nucleation or
percolation or $P_\text{f} < 0.01$ would not
happen before today, or have tachyonic modes in the zero-temperature vacuum state.
Further, we reject points of the
two-step model which have scalar masses below the muon threshold after the
transition, in order to allow for
sufficiently fast decays of the dark sector before BBN without the need of adding large bare
mass terms, as discussed in section~\ref{sec:flipflop}.
For details on the adopted \texttt{PTArcade} likelihood and the resulting
posterior distributions of the model parameters
see appendices~\ref{app:ptarcade} and \ref{app:triangles}, respectively.

We show the resulting best-fit regions in the right panel of figure~\ref{fig:summaryplot},
with $1\sigma$ and $2\sigma$ contours\footnote{%
\label{foot:sampling}
Due to the substantial numerical cost of the sampling, in particular for the Abelian dark
Higgs and flip–flop models, we deliberately terminated the sampling before full convergence
was reached. We note that these sampling challenges are directly related to the tuning issues discussed further
down. Extending the chains further would somewhat refine the detailed contour shapes, in particular, 
but not affect the location of the best-fit points. While the qualitative shape of the confidence 
regions shown in the figure should thus only be regarded as indicative, this has no impact on our 
qualitative conclusions -- and hence does not warrant additional computational effort.
}
relative to the respective best-fit points (represented by stars; see table~\ref{tab:benchmarks} 
in the appendix for the corresponding model parameter values). 
These best-fit points indeed all describe the data equally well,
as indicated by very similar likelihoods of
$\ln \mathcal{L} = -103.28$ ($-103.21$, $-103.42$) for the pink (blue, orange) star.

\begin{table}
  \begin{minipage}[t]{0.4\textwidth}
    \centering
    \textbf{Conformal model}\\[0.6em]
    \begin{tabular}{c|c}
      \toprule
      $ -v^{2}\frac{\dd^2}{\dd v^2}\ln \mathcal{L}$ & $ -g^{2}\frac{\dd^2}{\dd g^2}\ln \mathcal{L}$ \\
      \midrule
      & \\[-1.2em]
      
  $7\times10^{1}$ & $5\times10^{3}$  \\ 
    \bottomrule
    \end{tabular}\\[1em]  
  \end{minipage}
  \begin{minipage}[t]{0.565\textwidth}
    \centering
    \textbf{Abelian dark sector}\\[0.6em]
    \begin{tabular}{c|c|c}
      \toprule
      $ -v^{2}\frac{\dd^2}{\dd v^2}\ln \mathcal{L}$ & $ -\lambda^{2}\frac{\dd^2}{\dd \lambda^2}\ln \mathcal{L}$ & $ -\tilde g^{2}\frac{\dd^2}{\dd \tilde g^2}\ln \mathcal{L}$  \\
      \midrule
      && \\[-1.2em]
  $1\times10^{2}$ & $5\times10^{5}$  &   $4\times10^{8}$  \\ 
    \bottomrule
    \end{tabular}\\[1em]  
  \end{minipage}

  \vspace{2em}
  \centering
\textbf{Flip-flop} \\[0.6em]
  \begin{tabular}{c|c|c|c|c|c}
    \toprule
    $ -v^{2}\frac{\dd^2}{\dd v^2}\ln \mathcal{L}$ & $ -\lambda_0^{2}\frac{\dd^2}{\dd \lambda_0^2}\ln \mathcal{L}$ & $  -\lambda_{12}^{2}\frac{\dd^2}{\dd \lambda_{12}^2}\ln \mathcal{L}$ & $  -y^{2}\frac{\dd^2}{\dd y^2}\ln \mathcal{L}$  & $  -\lambda_{1}^{2}\frac{\dd^2}{\dd \lambda_{1}^2}\ln \mathcal{L}$ & $  -\gamma^{2}\frac{\dd^2}{\dd \gamma^2}\ln \mathcal{L}$ \\
    \midrule
    &&&&& \\[-1.2em]
$9\times10^{2}$ & $3\times10^{6}$  &  $2\times10^{7}$ &  $2\times10^{7}$ &  $5\times10^{7}$ &  $2\times10^{8}$ \\ 
    \bottomrule
\end{tabular}
\caption{The second derivative of $\ln \mathcal{L}$ with respect to the model parameters, for the best-fit
  points indicated with stars in the right panel of figure~\ref{fig:summaryplot} 
  (cf.~tables \ref{tab:benchmarks} and \ref{tab:benchmarks-observables} in appendix~\ref{app:triangles}).
  See text for a more detailed discussion of how these numbers relate to the relative parameter tuning that is 
  needed to explain the PTA signal in these models.}
\label{tab:tuning}
\end{table}

The PTA best-fit regions shown in the right panel of figure~\ref{fig:summaryplot} only overlap with 
the generic predictions (left panel) in the case of the conformal model, despite almost identical likelihoods,
and are significantly displaced for the other 
two models. This indicates that in the latter cases some tuning of the underlying model parameters is needed
in order to explain the PTA signal. 
In order to gain some intuition about how this statement may be quantified,
we consider the respective 
best-fit points and compute the second derivative of the likelihood in the direction of each model parameter. 
The result is shown in Table~\ref{tab:tuning}, confirming the general expectation from the discussion above:
the same change in $\ln \mathcal{L}$ requires a much larger change in the coupling $g$ 
of the conformal model compared to any of the coupling parameters appearing in the other two models.
Let us stress that while $\ba{-g^2 \tdd{}{g} \ln \mathcal{L}}^{-1/2}$ is a good indicator of the {\it relative} amount 
of tuning that is needed close to the respective best-fit points, in comparison between these models, 
it cannot be used to quantify this tuning in absolute terms. For the latter, one would rather be interested in the
same quantity for the {\it profiled} likelihood, which follows a $\chi^2$ distribution for any given parameter. 
An even more reliable estimator is the range of the respective model parameter within a given confidence interval,
as computed from the profiled likelihood. Due to the associated computational costs of fully profiling over all (other)
model parameters, we only performed this exercises for the conformal model. Profiling over $v$, we thus find
that $g$ can vary by roughly 15~\%  at the $2\,\sigma$ level -- indicating that, indeed, only a very
moderate amount of tuning is needed to fit the data in this case.
We note that the significant tuning required for the {\it other} models also explains the numerical challenges in 
precisely mapping out the confidence levels in these case, which we mentioned in footnote \ref{foot:sampling}.
It is worth remarking that the dependence on the relevant energy scale, which we parameterised in terms of the 
vev $v$, is quite weak in all models. This is due to the fact that changing $v$ merely shifts the peak frequency of the GW 
signal, which can be compensated by a change in $\beta/H$ to achieve a similar fit to the data.
This can also be seen by comparing the width of the posterior distributions in $v$ in
figures~\ref{fig:corner_thermal_barrier}, \ref{fig:corner_two_step} and \ref{fig:corner_conformal}, in
appendix~\ref{app:triangles} with the width of the 
posterior distributions for the couplings in the respective models.

\begin{figure}[t]
	\centering
	\vspace*{4cm}
  \includegraphics[width=\linewidth]{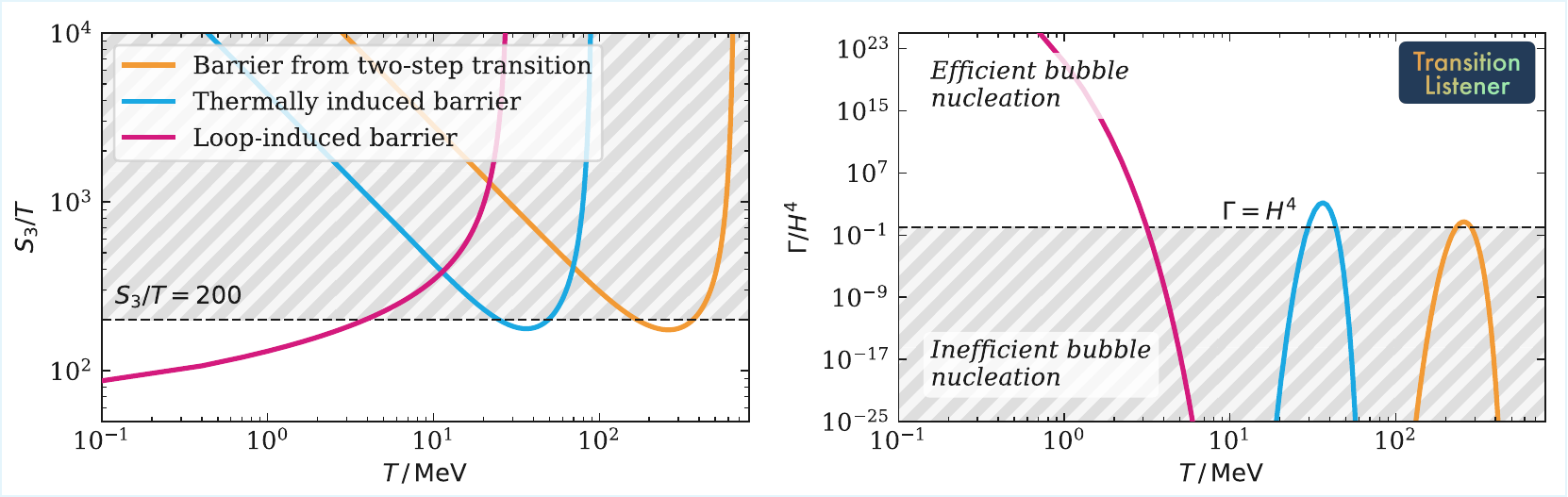}
  \caption{Comparison of the tunnelling action $S_3/T$ (left) and transition rate
    $\Gamma/H^4$ (right) as a function of temperature, for the three benchmark models defined
    in table~\ref{tab:benchmarks} (and indicated as stars in figure~\ref{fig:summaryplot}).}
	\label{fig:S3}
\end{figure}

We stress that the above discussion about the required tuning of model parameters does not strongly
depend on the specific parameterisation that we have chosen. Rather, it is a direct consequence of the
different physical behaviours of the models considered here, based on how the phase transition proceeds.
We illustrate this point in figure~\ref{fig:S3}, by showing the bounce action (left panel) and bubble nucleation
rate (right panel) for the three best-fit points discussed above.
As already anticipated from the discussion in section~\ref{sec:u1dark}, cf.~figure~\ref{fig:S3_U1}, the bounce
action of Abelian models with $\mu^2 \neq 0$ that fits the PTA data is U-shaped. This is expected because it is the only 
way of flattening $S_3/T$ sufficiently around $S_3/T \simeq 200$ to allow a slow PT,
and in particular avoiding the rapid  formation of bubbles significantly smaller than the original Hubble patch.
For the flip-flop model the situation is similar, though we generally find higher $T_\text{p}$ and lower 
$\alpha$ because less supercooling is necessary to produce a strong signal. Still, fine-tuning is needed 
to obtain small transition rates $\beta/H$, i.e.~an approximately flat $S_3/T$ just outside the hatched
area indicating inefficient bubble nucleation. In contrast, no  fine-tuning is needed in the conformal model
because, as a consequence of the logarithmic temperature dependence of the potential barrier, the transition 
automatically sets in at a low temperature. Here $S_3/T$ only acquires its minimum for $T\to0$
and is therefore always sufficiently flat around $S_3/T \simeq 200$ to guarantee a small $\beta/H$, 
without fine-tuning. 
We take the qualitatively different behaviour of the curves in figure~\ref{fig:S3} as indication
that the models we have studied here are indeed representative of three 
rather generic model classes that may be invoked as phase transition  interpretations of the PTA data. 

We complement the above discussion with a more detailed view of the underlying model
parameters in appendix~\ref{app:triangles}, where we show posterior distributions
in the form of triangle plots for each scenario after sampling over the PTA likelihood. These plots
make explicit which combinations of couplings and mass spectra are preferred once the PTA
fit requirement is imposed, and they provide a useful cross-check of the qualitative
tuning arguments discussed around table~\ref{tab:tuning}. In particular, they illustrate
that the conformal model achieves a good fit in a comparatively broad region of its
two-dimensional parameter space, whereas the Abelian Higgs and flip-flop scenarios only
populate narrow ridges or isolated patches of parameter space consistent with the data.
We also include posterior distributions of the derived particle masses after the phase transition
in these triangle plots, which we already referred to in section~\ref{sec:models} when discussing possible
decay channels and cosmological constraints. In general, we find that these distributions are much stronger
peaked than the corresponding prior distributions, indicating the exciting prospect 
that improved GW data might in principle 
be used to `measure' the masses of particles that remain inaccessible to direct laboratory searches. 
Since the couplings are in all cases relatively sharply peaked, the widths of the mass distributions mostly
reflect the widths of the posterior distributions of the vevs $v$. In accordance with what is visible
in figure~\ref{fig:S3}, the conformal model generally favours the smallest vev (peaking around $\sim200$\,MeV)
while the flip-flop model requires significantly larger vevs (peaking around $\sim1$\,GeV).
Correspondingly, both Abelian Higgs and conformal model typically feature scalar masses of 
$\mathcal{O}(10)$\,MeV, while the lighter scalar mass in the flip-flop model peaks at several hundred
MeV.  We note that this difference is at least partially driven by our hard cut of $m_{\phi_i}>2m_\mu$
in the latter case.

\section{Discussion of assumptions and caveats}
\label{sec:discussion}

In this section we discuss various assumptions
that went into our analysis and comment on possible caveats
and extensions of our work.
We start with effects that may (moderately) shift the
preferred regions in the $\alpha - \beta/H$ plane,
discuss other model classes than those considered above
and then turn to potentially critical issues.

\paragraph{Ultraslow transitions, PBHs, and GW
spectral templates:}
In figure~\ref{fig:model-independent} we indicated that
for $\beta/H \lesssim 10$ our results become less reliable,
because both the GW modelling and the transition dynamics
enter a regime not well covered by existing simulations.
First, the simulation-based scaling of the GW amplitude
with $\beta/H$ used in eq.~\eqref{eq:GWspec} has not been
validated for ultraslow transitions, and Hubble expansion
effects may lead to an additional suppression of the GW
signal~\cite{Lewicki:2025hxg}; if confirmed, this would
shift the preferred region towards smaller $\beta/H$.
Second, ultraslow transitions with almost Hubble-sized bubbles
have been argued to lead to the formation of a large abundance of 
primordial black holes (PBHs) and an additional, scalar-induced GW 
component~\cite{Kodama:1982sf, Lewicki:2023ioy, Gouttenoire:2023naa}. Recent works,
however, have questioned these claims, highlighting the importance
of a gauge-independent treatment of the curvature
perturbations~\cite{Franciolini:2025ztf, Wang:2026zvz}. Moreover, the common
assumption of a nucleation rate that is exponential in time
can fail in precisely the ultraslow regime
(cf.~figure~\ref{fig:S3}), making quantitative PBH
estimates model-dependent. Finally, our GW spectral template
is based on the latest ``Higgsless'' sound-wave
simulations~\cite{Jinno:2022mie, Caprini:2024gyk},
formally validated up to $\alpha \lesssim 0.5$. For
$\alpha \gg 1$ the spectrum is however expected to approach a single-broken
power law as in ref.~\cite{Lewicki:2022pdb} (used in the current
LISA pipeline~\cite{Caprini:2024hue}), with differences in the
low-frequency slope and amplitude scaling. Given the lack of
simulations interpolating between these regimes, we do not
attempt to fold these effects into the likelihood. Importantly,
any such refinements would predominantly affect the very small-$\beta/H$
region and thus mainly the (already strongly tuned) two-step case;
we therefore do not expect our main conclusions to change.

\paragraph{Energy budget time dependence:}
In our sound-wave prediction we follow the common choice of evaluating $\kappa_\text{sw}$
at percolation~\cite{Yamada:2025hfs}. For intermediate strengths
this can matter because $\alpha$ (and hence $\kappa_\text{sw}$)
is typically smaller at nucleation than at percolation, potentially
reducing the predicted GW amplitude by up to $\mathcal{O}(10^2)$
in extreme cases if $\kappa_\text{sw}$ is evaluated too early. These points would hence
not yield a good fit to the PTA signal and would need to be shifted
towards slightly higher $\alpha$. For $\alpha\gg 1$ this sensitivity
is mild since $\kappa_\text{sw}\approx 1$ at both nucleation and
percolation. A more faithful treatment would require capturing
the time dependence of the energy budget seen in recent
simulations~\cite{Caprini:2024gyk}. The only model where
this could affect the exact location of the best-fit region 
is again the two-step case,
where intermediate strengths in the PTA fit are common.

\paragraph{Alternative model mechanisms:}
We focussed on thermal transitions in weakly coupled scalar sectors,
but other mechanisms might modify or even bypass the
$\alpha - \beta/H$ correlations.
Non-thermal (quantum) phase transitions can yield parametrically
small transition speeds $\beta/H$ because the nucleation rate is
not temperature-dependent; explicit PTA-motivated examples
exist~\cite{Chatrchyan:2025wop}, with general expectations
discussed in~\cite{Freese:2022qrl}.
Seeded decay by topological defects (domain walls/cosmic strings)
can enhance nucleation and set the bubble scale by defect separations,
potentially allowing higher transition temperatures while keeping
PTA-scale peak frequencies~\cite{Blasi:2022woz, Blasi:2023rqi, Agrawal:2023cgp,
Chatrchyan:2025uar, Bai:2025qch}. In our flip-flop model, such effects could
in principle arise due to the first transition step generating
domain walls, which we however assumed to vanish sufficiently
quickly to not have a phenomenological impact.
Finally, strongly coupled confining sectors often feature only limited
supercooling, resulting in a suppressed GW signal. This expectation has been
emphasised recently~\cite{Agrawal:2025wvf}, with large-$N$ pure Yang--Mills being a
classic example~\cite{GarciaGarcia:2015fol}.  A notable exception are
near-conformal strongly coupled transitions, whose GW phenomenology we expect to be
qualitatively similar to the conformal scenario discussed in
section~\ref{sec:conformal}. Further work would be needed to robustly
establish this conclusion, cf.~ref.~\cite{Fujikura:2023lkn} for a first step in this  direction.

\paragraph{Homogenisation before BBN:}
It has been argued that strong first-order phase transitions
can produce large-scale inhomogeneities that may survive until BBN
and spoil the successful nucleosynthesis predictions~\cite{Bagherian:2025puf}.
In our MeV-scale setting, the concern could be even more acute due
to the shorter time available for homogenisation. However, existing
estimates rely on simplified homogenisation dynamics (i.e.,~baryon
diffusion as the dominant effect), whereas strong FOPTs generically
also generate bulk flows and turbulence that may accelerate homogenisation.
Given these uncertainties, we do not attempt to include homogenisation
constraints here. It should be noted, however, that these effects
would predominantly affect the very strong transition regime relevant
for the conformal model, and we therefore cannot exclude that they
may impact our conclusions in this case.

\medskip
In summary, even though various assumptions and caveats in principle
apply to our analysis, we do not expect them to significantly affect our main
conclusions. 
Both the selected model classes and our treatment of the generated GW
spectrum, in other words, can thus be understood as
representative of rather generic possibilities to explain the PTA
signal in terms of a first-order
phase transition in a dark sector. 
Still, future work is needed to address
some of the discussed issues in more detail, especially in the case
of ultraslow transitions, in order to fully exploit the potential of PTA
observations to probe the dynamics of first-order phase transitions 
in the early universe.

\section{Conclusions}
\label{sec:conclusion}

The recent observation of a common red noise signal by various PTA collaborations
offers a new probe of strong dynamics in the early universe through the associated
stochastic gravitational wave background. A particularly interesting possibility
is that this signal originates from a dark sector first-order phase transition
at temperatures around the MeV scale. In this work we have scrutinised the
viability of this explanation.
In the first part (section~\ref{sec:model-independent}), we therefore started by performing a model-independent
analysis of the 15\,yr PTA data to infer the preferred regions in the parameter space spanned
by transition strength $\alpha$, speed $\beta/H$ and reheating
temperature $T_\text{reh}$ of the phase transition. 
We thereby updated earlier results based on previous data releases, cf.~figure~\ref{fig:model-independent}, confirming 
that a phase transition  interpretation of the data requires a transition that is both rather strong
and slow.

In the remainder of our work, we then studied three representative model classes 
that can give rise to a first-order phase transition consistent with these observations.
These model classes are based on \textit{(i)} a thermally induced barrier in a higgsed
Abelian dark sector, \textit{(ii)} a two-step `flip-flop' transition in a model with
two scalar singlets, and \textit{(iii)} a loop-induced barrier in an Abelian dark sector that
is classically conformally invariant. Among these, we conclude that only the conformal model can
explain the PTA signal without significant tuning requirements for the model parameters.
We stress that this result is rather generic and can be traced back to the
different physical behaviours of the models, based on how the phase transition proceeds 
(cf.~figures~\ref{fig:summaryplot} and~\ref{fig:S3}):

\begin{itemize}
\item In the case of a thermally induced barrier, a moderate amount of
supercooling is needed to fit the PTA data and the
nucleation rate must be significantly tuned to achieve slow transitions  
$20 \lesssim \beta/H \lesssim 100$.
The intrinsic reason for this tuning is 
that the bubble nucleation rate has to peak at sufficiently low temperatures
to achieve strong supercooling, while at the same time being sufficiently
flat around the percolation temperature to guarantee a slow transition.

\item Somewhat less supercooling, $\alpha \sim 1$, is required in the case of a two-step
transition; in this case, however, the transition speed must be tuned down to
$\beta/H \lesssim 10$ in order to explain the PTA signal, stretching the validity of the employed
GW spectral templates. The large amount of tuning is again due to 
the need of flattening the bubble nucleation rate around the percolation temperature. 
Crucially,  the presence of an additional
field direction in which tunnelling can proceed typically suppresses the
bounce action and enhances the bubble nucleation rate.

\item In contrast, the conformal model naturally leads to strong supercooling and
slow transitions, such that for $10^1 \lesssim \alpha \lesssim 10^6$ and
$30 \lesssim \beta/H \lesssim 100$ the PTA signal can be explained without
any significant tuning of the model parameters. The reason for this
behaviour is the weak temperature dependence of the potential barrier,
which automatically leads to a transition at low temperatures after strong
supercooling: Here,  the bounce action $S_3/T$ only acquires its minimum for $T\to0$
and is therefore always sufficiently flat around  percolation 
to guarantee a slow transition.
\end{itemize}

In our analysis we discussed various ways in which additional constraints from
BBN and CMB observations, as well as collider searches for light new particles,
may be avoided. Most importantly, we find that while the GWs themselves result 
in a negligible contribution to $\Delta N_\text{eff}$, the same does not hold for the
additional light degrees of freedom in the dark sector after the transition. 
Generally speaking, those contributions to $\Delta N_\text{eff}$ can be prevented by allowing 
scalar or gauge boson decays into SM particles before BBN, or by introducing additional 
dark sector states that help in thermalising the dark sector with the SM. 
Whether these options are viable is a model-dependent question that we 
comment on in some detail for each of the scenarios studied here.
We also discuss various 
potential caveats and possible further refinements of our analysis,
but argue that none of those affects our main conclusion. All three model
classes considered here can thus indeed be seen as representative of rather generic possibilities
to explain the PTA signal in terms of a first-order phase transition in a dark sector.
In particular, it appears challenging to address the tuning issues without a potential barrier
that is only radiatively induced, though it should be noted that in that case complementary 
constraints are more difficult to avoid.

Further work will be beneficial to address some of the discussed assumptions
and caveats in more detail once even more precise PTA data become available.
This is expected to happen already in the near future, with the third IPTA data 
release~\cite{InternationalPulsarTimingArray:2023mzf}
as well as in upcoming data releases of the individual PTA collaborations.
After having confirmed the gravitational wave nature of the
observed common red noise signal, in particular, the most important next steps will be 
to determine its spectral shape more precisely, and whether the signal is 
isotropic~\cite{Konstandin:2024fyo, Konstandin:2025ifn}. This will make it possible to 
better discriminate between different
sources, including first-order phase transitions and supermassive black hole
binaries. Should a first-order phase transition remain a plausible option, upcoming data will also
shrink the preferred regions in the generic $\alpha - \beta/H - T_\text{reh}$
parameter space, which in turn will help to further discriminate between different model classes.
Our work provides a first step in this direction, by identifying already at this stage which model classes 
appear to most naturally explain the observed signal.

\acknowledgments We thank Aleksandr Chatrchyan, Majid Ekhterachian, Felix Kahlhoefer,
Maciek Kierkla, William Lamb, and Alberto Roper Pol for useful discussions. TB and CT
gratefully acknowledge support from the Alexander von Humboldt Foundation -- via a
Friedrich Wilhelm Bessel Award (TB) and a Feodor Lynen Research Fellowship (CT) -- as well
as through a FRIPRO grant of the Norwegian Research Council (project ID 353561
‘DarkTurns’). 
This work is also funded by the Deutsche Forschungsgemeinschaft (DFG) through Germany's Excellence Strategy --- EXC 2121 ``Quantum Universe'' --- 390833306 and through
Grant No. 396021762 – TRR 257.

\newpage
\appendix	
\addtocontents{toc}{\protect\setcounter{tocdepth}{1}}

\section{Details on the PTArcade likelihood}
\label{app:ptarcade}

The PTArcade likelihood $\ln \mathcal{L}_\text{PTArcade}$ is only valid for 
$\ln \mathcal{L}_\text{PTArcade}\geq \tau\approx -110$ because it features unphysical 
plateaus below this value. For smaller values of $\ln \mathcal{L}_\text{PTArcade}$, 
we therefore use an approximate likelihood $\mathcal{L}_\text{approx}$ that penalises 
parameter points which are too far away from the region of interest. This approximate 
likelihood is constructed to become maximal when the peak frequency and amplitude 
of the GW signal correspond to the region preferred by PTA 
data, cf.~right panel of figure~\ref{fig:model-independent}. We thus use a combined likelihood 
$\mathcal{L}_\text{PTA}$ with
\begin{align}
\ln \mathcal{L}_\text{PTA} &= \ln \mathcal{L}_\text{PTArcade} + \Theta( \tau -\ln \mathcal{L}_\text{PTArcade}) \times \bb{\tau -\ln \mathcal{L}_\text{PTArcade} + \ln \mathcal{L}_\text{approx}}\,, \nonumber \\ 
\ln \mathcal{L}_\text{approx} &=  - (\log_{10} (h^2 \Omega^\text{peak}_\text{GW}) + 8.13 )^2 - (\log_{10} (f^\text{peak}_\text{GW} / \text{Hz}) + 7.70 )^2 \, ,
\end{align}
which ensures a smooth transition between both likelihoods. Using $\mathcal{L}_\text{PTA}$ 
leads to a quicker convergence of the sampling algorithm than using $\mathcal{L}_\text{PTArcade}$, 
without compromising accuracy. 
The origin of the plateaus in the PTArcade likelihood is found in the \texttt{ceffyl} source 
code\footnote{\url{https://github.com/astrolamb/ceffyl/blob/32f5b43cd8d6d1b3ee262c2a11f01eec64df34d5/ceffyl/Ceffyl.py\#L505}},
which sets the  likelihood of a given element of the Bayesian spectrogram (`violin')
to a constant value if the GW amplitude at that frequency is outside 
the prior range of the violin. Note further that the threshold parameter has a value of $\tau \approx -110$ only 
in recent versions of \texttt{ceffyl} (starting from a normalisation fix introduced in September~2023) 
and consequently in \texttt{PTArcade} versions that correctly depend on these releases (in particular 
\texttt{PTArcade}~$\geq$~v1.1.5). In earlier \texttt{PTArcade} versions (e.g.\ v1.1.1), one instead needs 
to use $\tau = -55$, due to an undocumented \texttt{tempo2} installation issue, which forced the use of an older
\texttt{ceffyl} release. The installation issue has been resolved in later \texttt{PTArcade} releases, such
that the more recent versions of \texttt{PTArcade} can now be used for a consistent Bayesian model
comparison. The normalisation of the likelihood 
has no impact on parameter estimation or frequentist likelihood analyses, as performed in this work.

\section{Effective potential for  the flip-flop model}
\label{app:flipflop}

In this appendix we provide further details about the effective potential computation for the
two-step transition model introduced in section~\ref{sec:flipflop}. The tree-level potential in
eq.~\eqref{eq:flipflop-potential} has four minima located at
\begin{align}
  \left\langle \phi \right\rangle_{1,2} =  \begin{pmatrix} 
 0\\
    \pm \,v_2 
 \end{pmatrix}
  \equiv  \begin{pmatrix} 
 0\\
    \pm \,{v}/{\gamma} 
 \end{pmatrix}
 \quad \text{and} \quad
 \left\langle \phi \right\rangle_{3,4} =  \begin{pmatrix} 
 \pm \,v_1 \\
    0
 \end{pmatrix}
  \equiv 
   \begin{pmatrix} 
\pm \sqrt{1 + {\lambda_1}/{\lambda_{0}}} \,v \\
0
 \end{pmatrix} \, ,
\end{align}
at which the tree-level potential  becomes
\begin{align}
 V_{\mathrm{tree}}(0, \pm v_{2}) = 0
  \qquad \text{and} \qquad
  V_{\mathrm{tree}}(\pm v_{1}, 0) = - \frac{\lambda_{1}}{4} \left( 2 +
  \frac{\lambda_1}{\lambda_{0}} \right) v^{4}
 \, .
\end{align}
The background field-dependent mass matrix for the scalars reads
\begin{align}\label{eq:flipflop-mass-matrix}
  &\mathcal{M}^2 \left( \phi_{1}, \phi_{2} \right) = \notag\\
&~~~\begin{pmatrix} 
  -v^2 (\lambda_{0} + \lambda_{1}) + 3 \lambda_0 \phi_{1}^2+ \phi_{2}^2
  \left(\gamma ^2 \lambda_0+ \lambda_{12}\right)
  & 2 \phi_{1} \phi_{2} \left(\gamma ^2 \lambda_{0}+ \lambda_{12}\right) \\
  2 \phi_{1} \phi_2 \left(\gamma ^2 \lambda_{0}+ \lambda_{12}\right)
  & -\gamma ^2 \lambda_{0} v^2+ \phi_{1}^2 \left(\gamma ^2 \lambda_{0}+
  \lambda_{12} \right)+3 \gamma ^4
    \lambda_{0} \phi_{2}^2 \\
\end{pmatrix}\,.
\end{align}
Since $\lambda_0, \lambda_1 > 0$ and hence
$V_\text{tree}(\langle \phi \rangle_{3,4}) < V_\text{tree}(\langle \phi \rangle_{1,2})$,
the global minimum lies on the $\phi_{1}$-axis where
the off-diagonal elements of $\mathcal{M}$ vanish. Ignoring the impact of domain walls and
in order to simplify the analysis (see discussion in section~\ref{sec:flipflop}),
we choose $\left\langle \phi \right\rangle_3 = (+v_{1}, 0)$ as the vev in the
broken phase after the transition.

In order to 
be able to use the vev as an input parameter for our
numerical scan, we employ an on-shell-like scheme for renormalising the 1-loop effective
potential, as introduced in ref.~\cite{Basler:2018cwe} for the study of phase transitions in
multi-Higgs models. In doing so,
the vev and the masses of the two dark singlet scalars are kept at their tree-level values
after including the Coleman-Weinberg contribution to the potential. The counterterm potential
therefore has the form
\begin{align}\label{eq:counterterms_flipflop}
  V_{\mathrm{ct}} =  - \frac{\delta\mu_1^2}{2} \phi_1^{2}
  - \frac{\delta\mu_2^2}{2} \phi_{2}^{2} + \frac{\delta\lambda_{1}}{4}\phi_{1}^{4}  \,,
\end{align}
where the counterterms are determined by the renormalisation conditions
\begin{align}
  \label{eq:counterterm-conditions-flipflop}
 \delta \lambda_{1} &= \frac{1}{2 v_1^3} \frac{\partial V_{\mathrm{CW}}}{\partial \phi_{1}}\Big |_{(v_{1}, 0)} -
            \frac{1}{2 v^2_{1}} \frac{\partial^2V_{\mathrm{CW}}}{\partial\phi^2_{2}}\Big |_{(v_{1}, 0)} \, , \\
 \delta\mu_{1}^{2} &= \frac{3}{2 v_{1}} \frac{\partial V_{\mathrm{CW}}}{\partial\phi_{2}}\Big |_{(v_{1}, 0)} -
                      \frac{1}{2} \frac{\partial^2V_{\mathrm{CW}}}{\partial\phi^2_{2}}\Big |_{(v_{1}, 0)}  \, , \qquad \text{and} \\ 
 \delta\mu_{2}^2 &= \frac{\partial^2V_{\mathrm{CW}}}{\partial\phi^2_{2}}\Big |_{(v_{1}, 0)} \, . 
\end{align}
To account for the resummation of daisy diagrams, we employ the Arnold-Espinosa
method and add the thermal mass corrections~\cite{Arnold:1992rz}
\begin{align}
  \label{eq:hard-thermal-masses-flipflop}
  \Pi_{\phi_{1}}(T) &= \frac{T^2}{24}  \left( y^2 + 2 (3+ \gamma^2)\lambda_{0} +
  \lambda_{12} \right) \quad \text{and} \quad \Pi_{\phi_{2}}(T)
  = \frac{T^2}{12} \left( \gamma^2\lambda_{0}  + 3 \gamma^{4} \lambda_{0} +\lambda_{12}\right)\
\end{align}
to the diagonal elements of the mass matrix in eq.~\eqref{eq:flipflop-mass-matrix} in the
daisy contribution to the thermal potential, cf.~eq.~\eqref{eq:effectivepotential}. We obtained
these hard thermal masses by computing the thermal self-energy matrix
$\Pi(T) = \partial^2 V_{T}/(\partial\phi_{i}\partial\phi_{j})$ to leading order in temperature
and couplings.

\section{Model parameter posterior distributions and benchmark points}
\label{app:triangles}

In this appendix, we present posterior distributions (`triangle plots')
of the model parameters,
and the derived masses of the dark sector states after the phase transition,
for the three models considered in the main text. In each case, we show
the original random samples in grey, along with the preferred regions
after sampling over the PTA likelihood in colour. We  summarise our
benchmark points, 
corresponding to the best-fit points of the latter scans, 
and their corresponding phase transition parameters in
tables~\ref{tab:benchmarks} and~\ref{tab:benchmarks-observables}, respectively.
We checked that the bubble wall velocity for all three models is $v_\text{w} \simeq 1$ 
when using the LTE approximation for the friction~\cite{Ai:2023see}.

\begin{table}[t!]
\centering
\begin{tabular}{lll}
\toprule
Model & Parameter & Value \\
\midrule
\multirow{3}{*}{Abelian Higgs $U(1)'$} & $\tilde{g}$ & 2.7829 \\ 
                                                & $\lambda$ &  0.0406 \\ 

                                                & $v / {\rm GeV}$ & 0.24 \\ 
\midrule
\multirow{2}{*}{Conformal $U(1)'$} & $g$ & 0.692 \\ 
                                   & $v / {\rm GeV}$ & 0.14 \\ 
\midrule
                                                               
  \multirow{6}{*}{Flip-flop} & $\lambda_0$          & 0.02432 \\ 
                             & $\lambda_1$          & 0.014524 \\ 
                             & $\lambda_{12}$         & 0.023401 \\ 
                             & $v / {\rm GeV}$ & 0.927 \\ 
                             & $y$             & 1.0 \\
                             & $\gamma$             & 1.0\\
\bottomrule
\end{tabular}
\caption{Benchmark parameter values for the three example models highlighted in the main text. 
We quote all significant digits obtained in the numerical analysis up to the degree of accuracy
that is relevant for reproducing the PTA fit (cf.~table \ref{tab:tuning} in the main text). 
For the derived model parameters, in particular the particle masses after the phase transition, see
figures~\ref{fig:corner_thermal_barrier}, \ref{fig:corner_two_step}, and \ref{fig:corner_conformal}.} 
\label{tab:benchmarks}
\end{table}

\begin{table}[t!]
\centering
\begin{tabular}{lcccccccc}
\toprule
Benchmark point & $\alpha$ & $\beta/H$ & $RH_*$ & $T_\text{nuc}$ & $T_\text{p}$ & $T_\text{reh}$ & $\ln \mathcal{L}_\text{PTArcade}$ \\
 & & & & MeV & MeV & MeV& \\
\midrule
  Abelian dark Higgs
                & 1.74 & 22.8 & 0.19 & 43.6 & 34.7 & 46.0 & $-103.21$ \\ 
Conformal $U(1)'$ & 821 & 57.9
& 0.08
& 3.16 
& 2.58 & 14.5 & $-103.28$ \\ 
Flip-flop & 0.70 & 5.5 & 0.80 & 285 & 173 & 192 & $-103.42$ \\
\bottomrule
\end{tabular}
\caption{Phase transition parameters for the three benchmark points given in
  table~\ref{tab:benchmarks}. The last column shows the log-likelihood values for the
  NANOGrav 15\,yr data obtained using \texttt{PTArcade} v1.1.5.}
\label{tab:benchmarks-observables}
\end{table}

\begin{figure}
	\centering
  \includegraphics[width=0.9\linewidth]{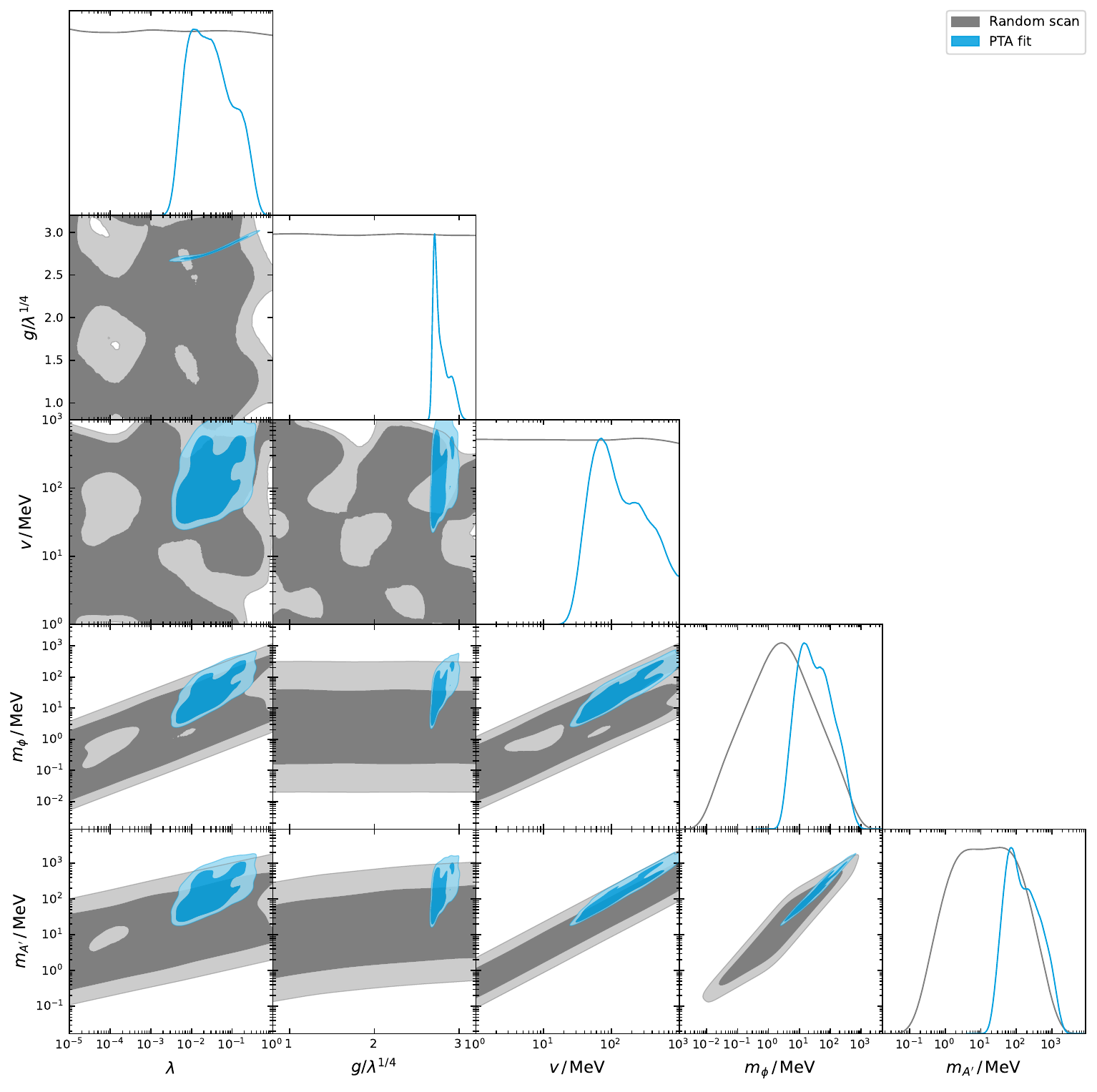}
	\caption{Triangle plot showing the preferred regions of the Abelian dark Higgs model
  after sampling over the PTA likelihood. The random samples are shown in grey.}
	\label{fig:corner_thermal_barrier}
\end{figure}

Figure~\ref{fig:corner_thermal_barrier} shows the results of our random and \texttt{ultranest} scans over
the Abelian dark Higgs model parameter space. The posterior of the coupling combination $\tilde{g}=g / \lambda^{1/4}$
has a broad peak at $\tilde{g} \simeq 2.5 - 3$, which corresponds to the requirement of 
a sufficiently strong phase transition. For larger $\tilde{g}$, the transition becomes too strong and the 
GW amplitude overshoots the PTA data, eventually also leading to percolation problems due to excessive supercooling. 
Conversely, for smaller $\tilde{g}$ the transition becomes too weak to explain the PTA signal. The quartic coupling 
$\lambda$ also has a broad distribution between $10^{-3}$ and $1$; note however that for each
given $\lambda$, the value of $\tilde{g}$ to obtain a good PTA fit needs to be tuned up to the fourth significant digit, 
cf.~table~\ref{tab:tuning}.
Correspondingly, the posterior distribution of the gauge coupling $g = \tilde{g} \lambda^{1/4}$
(not shown in the plot) also has a broad distribution ranging from $0.5$ to $2.5$.
The resulting mass distributions of the dark Higgs and dark gauge boson peak around
$m_\phi \sim 10 \, \text{MeV}$ and $m_{A'} \sim 100 \, \text{MeV}$, each with
tails spanning one order of magnitude in mass. The vev $v$ is  the least
constrained parameter, peaking at $\sim100$\,MeV. 
The dominant effect of a specific vev on the GW signal is to set the scale of the peak frequency, 
thus resulting only in a very mild degree of tuning of this parameter for a fixed set of couplings.
Correspondingly, the posterior distribution of $v$ is rather flat, consistent with the discussion
in section~\ref{sec:u1dark}.
We note that it is mostly the left shoulder
of the distribution in $v$ that is directly driven by data, coming from the requirement that the peak of the spectrum 
must not be located at frequencies $f \ll 10^{-8}$\,Hz, cf.~figure~\ref{fig:model-independent}. The fact that larger
values of $v$ are disfavoured, in contrast, is a consequence of an increasing difficulty to achieve 
sufficiently low $\beta/H$ (needed to maintain the required overall amplitude). As long as the data themselves
do not show clear evidence for a peak structure, let alone its exact location, 
the upper bound on the $v$ distribution is thus more a model 
limitation than something that can be inferred from the data.
Note that the cutoff of the posterior distribution at the upper
prior boundary also contributes to the mild cut-off in the mass distributions visible in $m_{A^\prime}$.

\begin{figure}
	\centering
  \includegraphics[width=0.9\linewidth]{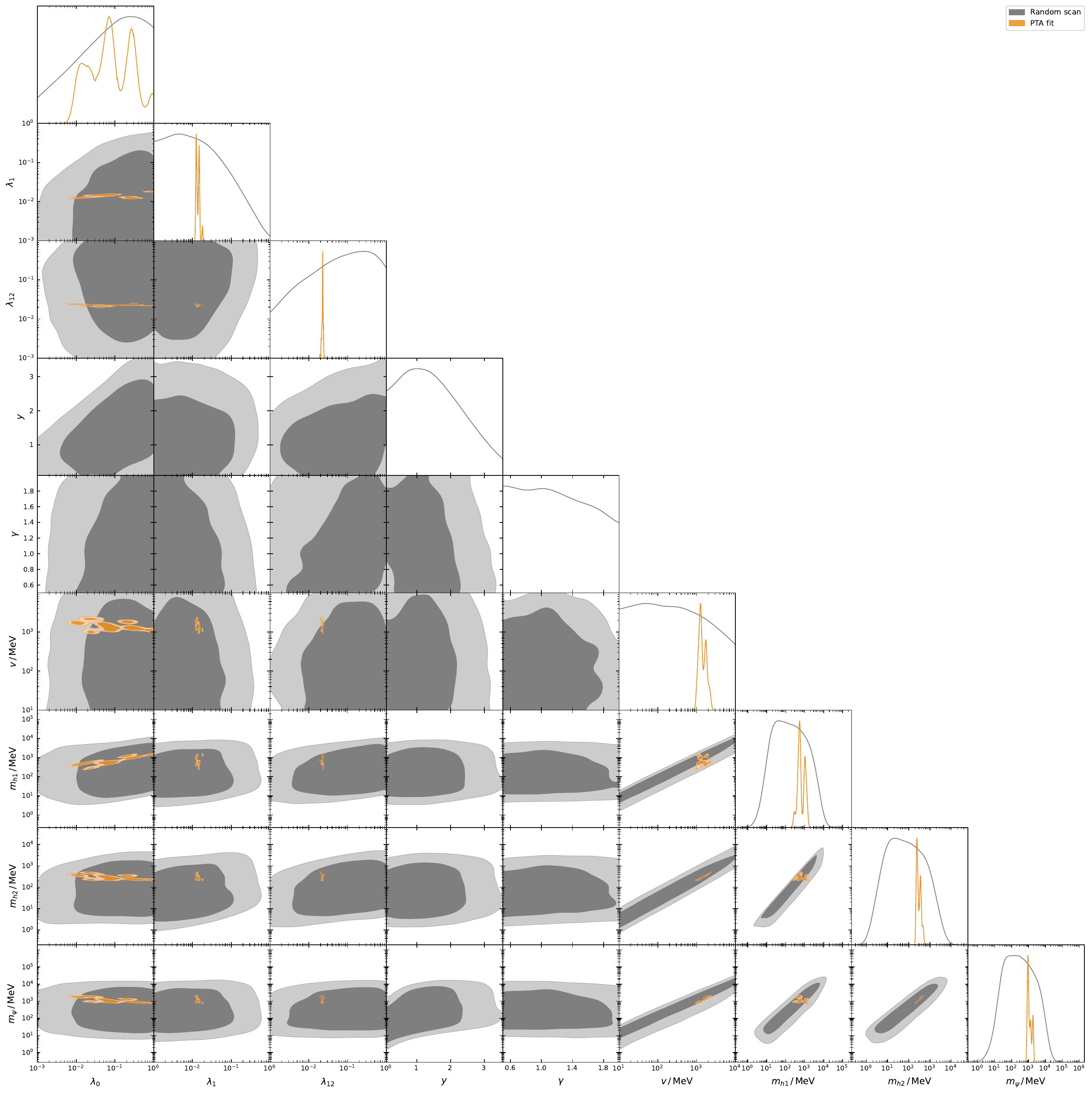}
	\caption{Triangle plot showing the preferred regions of the flip-flop model
  after sampling over the PTA likelihood and imposing the mass cuts on the dark singlets. The random samples are shown in grey.
  Note that the couplings $y$ and $\gamma$ have been fixed to unity for the PTA fit
  (indicated by missing posterior distributions in the respective panels) and that the posteriors in
  the $\lambda_0$ direction
  are rather poorly sampled due to numerical challenges in achieving a sufficiently slow phase transition.}
	\label{fig:corner_two_step}
\end{figure}

Next, figure~\ref{fig:corner_two_step} shows our results for the flip-flop model with a
two-step phase transition. Here the vev $v$ is again the least constrained parameter,
centred around $v\sim1 \, \text{GeV}$ and thus at a considerably larger value than
for the other models (see also figure~\ref{fig:S3}). Note that we imposed $y = \gamma = 1$
in order to simplify the numerical analysis when fitting the PTA signal;
relaxing this assumption would likely
broaden the preferred range of $v$ even further. The quartic couplings
$\lambda_{1}$ and $\lambda_{12}$ are as discussed in the main text highly tuned to
$\mathcal{O}(10^{-2})$ values,
featuring very narrow posterior distributions.
The quartic coupling $\lambda_{0}$, dictating the overall shape of
the potential, is larger and generally less constrained
(though still strongly peaked at its central value -- which is however
likely a sampling effect, see above). The
total number of sampled points that yield a good fit to the PTA data,
particularly visible in the panels with $\lambda_0$ on the axes, is
significantly smaller than for the other two models, indicating the
numerical challenges in identifying parameter points yielding sufficiently
slow phase transition in this model. The imposed mass cuts of both dark
scalars to be above the muon threshold after the transition,
$m_{\phi_i} > 211 \, \text{MeV}$, is clearly
visible in their mass distributions, which peak between $300 \, \text{MeV}$ and
$1 \, \text{GeV}$, with slightly larger $m_{\phi_1}$ than $m_{\phi_2}$
in general. The fermion is typically the heaviest dark sector state,
with a mass sharply peaking at around $m_\psi \simeq 1 \, \text{GeV}$,
thus safely avoiding BBN constraints via its
Boltzmann suppression prior to the decay of the dark scalars.

\begin{figure}
	\centering
  \includegraphics[width=0.9\linewidth]{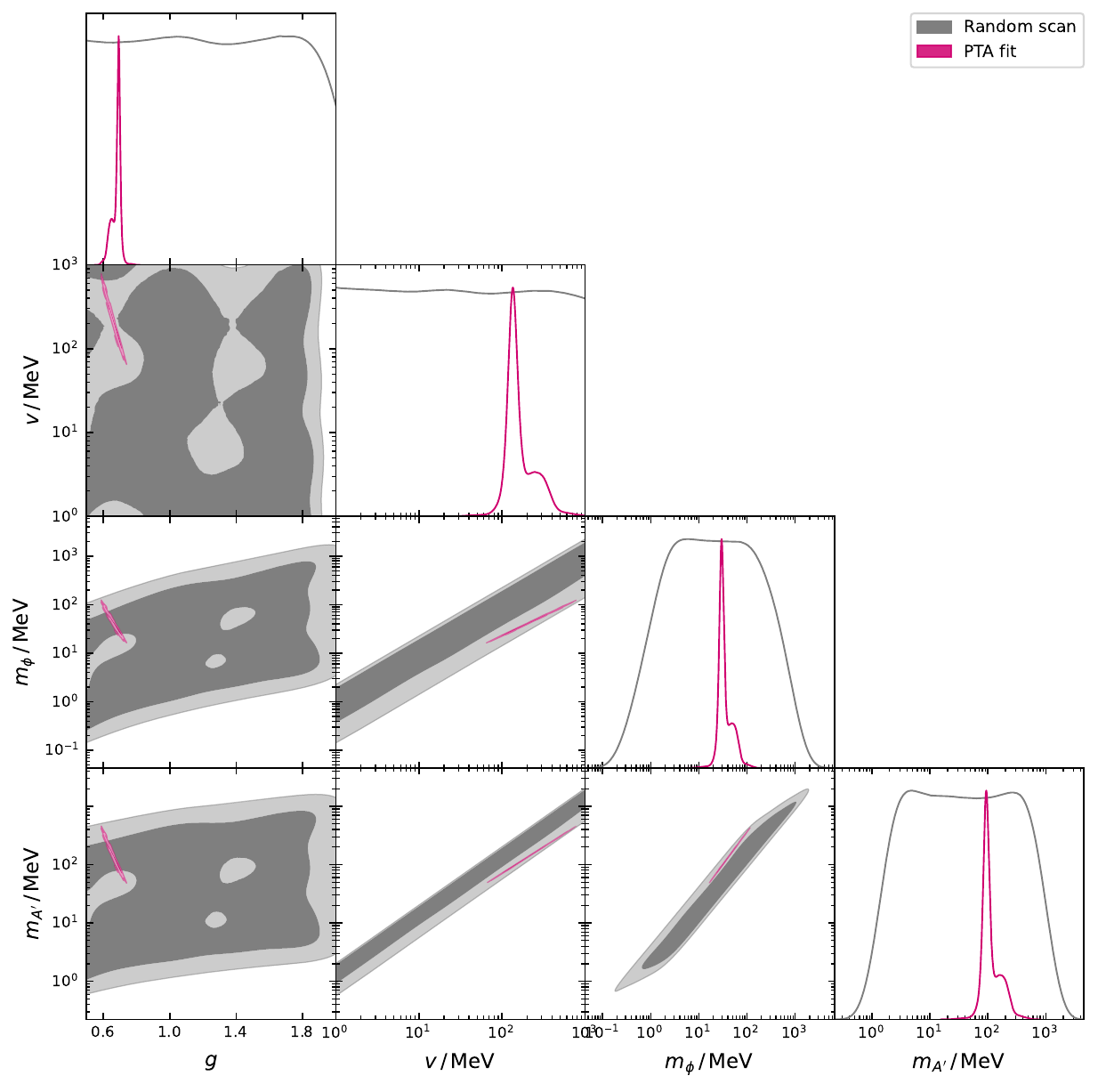}
	\caption{Triangle plot showing the preferred regions of the conformal model
  after sampling over the PTA likelihood. The random samples are shown in grey.}
	\label{fig:corner_conformal}
\end{figure}

In figure~\ref{fig:corner_conformal}, finally, we display the results for the conformal model.
We find that the preferred value of the dark Higgs vev after the transition
lies around  $v \simeq 200 \, \text{MeV}$, though at $2\sigma$ a range of
$80\, \text{MeV}\lesssim v\lesssim400\, \text{MeV}$ is allowed.
The dark gauge coupling is more strongly constrained, peaking around $g \simeq 0.6 - 0.7$
in accordance with the results presented in ref.~\cite{Balan:2025uke}. This implies that the
resulting masses of the dark Higgs and dark gauge boson typically lie in the range of a few
tens of MeV and around $100 \, \text{MeV}$, respectively.

\bibliographystyle{JHEP_improved}
\bibliography{bibliography}

\end{document}